\documentclass[aps, prx, reprint, twocolumn,superscriptaddress,floatfix]{revtex4-1}

\pdfoutput=1
\usepackage{amsmath}
\usepackage{verbatim}
\usepackage{graphicx}
\usepackage{color}
\usepackage{ragged2e}
\usepackage{wrapfig}
\usepackage{mathtools}
\usepackage{bbold}
\usepackage{blkarray}
\usepackage[inline]{enumitem}
\usepackage{latexsym}
\usepackage{array}
\usepackage{colortbl}
\usepackage{soul}
\usepackage{hyperref}
\hypersetup{colorlinks=true,        
    linkcolor=black,         
    citecolor=black,        
    filecolor=black,      
    urlcolor=black }

\newcommand{\ket}[1]{|#1\rangle} 
\newcommand{\bra}[1]{\langle#1|} 
\newcommand{\braket}[2]{\langle #1 |  #2  \rangle} 
\newcommand{\expval}[1]{\langle#1\rangle} 
\newcommand{\luca}{\color{green}}
\newcommand{\jan}{\color{blue}}

\begin{document}


\title{Towards simulating 2D effects in lattice gauge theories on a quantum computer}

\author{Danny Paulson}
\thanks{These authors contributed equally.}
\author{Luca Dellantonio}
\thanks{These authors contributed equally.}
\author{Jan F. Haase}
\thanks{These authors contributed equally.}
\affiliation{Institute for Quantum Computing, University of Waterloo, Waterloo, Ontario, Canada, N2L 3G1}
\affiliation{Department of Physics \& Astronomy, University of Waterloo, Waterloo, Ontario, Canada, N2L 3G1}
\author{Alessio Celi}
\affiliation{Departament de F\'isica, Universitat Aut\`onoma de Barcelona, E-08193 Bellaterra, Spain}
\affiliation{Center of Quantum Physics, University of Innsbruck, A-6020, Innsbruck, Austria}
\affiliation{Institute for Quantum Optics and Quantum Information of the Austrian Academy of Sciences, A-6020 Innsbruck, Austria}
\author{Angus Kan}
\affiliation{Institute for Quantum Computing, University of Waterloo, Waterloo, Ontario, Canada, N2L 3G1}
\affiliation{Department of Physics \& Astronomy, University of Waterloo, Waterloo, Ontario, Canada, N2L 3G1}
\author{Andrew Jena}
\affiliation{Institute for Quantum Computing, University of Waterloo, Waterloo, Ontario, Canada, N2L 3G1}
\affiliation{Department of Combinatorics \& Optimization, University of Waterloo, Waterloo, Ontario, Canada, N2L 3G1}
\author{Christian Kokail}
\author{Rick van Bijnen}
\affiliation{Center of Quantum Physics, University of Innsbruck, A-6020, Innsbruck, Austria}
\affiliation{Institute for Quantum Optics and Quantum Information of the Austrian Academy of Sciences, A-6020 Innsbruck, Austria}
\author{Karl Jansen}
\affiliation{NIC, DESY Zeuthen, Platanenallee 6, 15738 Zeuthen, Germany}
\author{Peter Zoller}
\affiliation{Center of Quantum Physics, University of Innsbruck, A-6020, Innsbruck, Austria}
\affiliation{Institute for Quantum Optics and Quantum Information of the Austrian Academy of Sciences, A-6020 Innsbruck, Austria}
\author{Christine A. Muschik}
\email{christine.muschik@uwaterloo.ca}
\affiliation{Institute for Quantum Computing, University of Waterloo, Waterloo, Ontario, Canada, N2L 3G1}
\affiliation{Department of Physics \& Astronomy, University of Waterloo, Waterloo, Ontario, Canada, N2L 3G1}
\affiliation{Perimeter Institute for Theoretical Physics, Waterloo, Ontario N2L 2Y5, Canada}

\date{\today}

\begin{abstract}
\noindent 
Quantum computing is in its greatest upswing, with so-called Noisy-Intermediate-Scale-Quantum
devices heralding the computational power to be expected in the near future. While the field is
progressing towards quantum advantage, quantum computers already have the potential to
tackle classically intractable problems. Here, we consider gauge theories describing fundamental particle interactions.
On the way to their full-fledged quantum simulations, the challenge of limited resources on near-term quantum devices has to be overcome. We propose an experimental quantum simulation scheme to study ground state properties in two-dimensional quantum electrodynamics (2D QED) using existing quantum technology. Our protocols can be adapted to larger lattices and offer the perspective to connect the lattice simulation to low energy observable quantities, e.g. the hadron spectrum, in the continuum theory. By including both dynamical matter and a non-minimal gauge field truncation, we provide the novel opportunity to observe 2D effects on present-day quantum hardware. More specifically, we present two Variational Quantum Eigensolver (VQE) based protocols for the study of magnetic field effects, and for taking an important first step towards computing the running coupling of QED. For both instances, we include variational quantum circuits for qubit-based hardware. We simulate the proposed VQE experiments classically to calculate the required measurement budget under realistic conditions. While this feasibility analysis is done for trapped ions, our approach can be directly adapted to other platforms. The techniques presented here, combined with advancements in quantum hardware, pave the way for reaching beyond the capabilities of classical simulations.
\end{abstract}

\maketitle


\section{Introduction}
\label{sec:introduction}
\noindent Quantum computers are at the edge of pushing our computational capabilities beyond the boundaries set by classical machines \cite{arute2019quantum,boixo2018characterizing,lund2017quantum}. One of the fields where quantum computers are particularly promising is the simulation of gauge theories, which describe the interactions of elementary particles. While current numerical simulations led to several breakthroughs \cite{aoki2020flag, aoki2017review, PhysRevD.100.114501}, they are ultimately restricted in their predictive
capabilities. On one side, limitations originate from the inherent difficulty faced by classical computers
in simulating quantum properties. On the other, the sign problem \cite{Troyer:2004ge, Gattringer:2016kco} affects simulations of both equilibrium (e.g., phase diagrams), and non-equilibrium (e.g., real-time dynamics) physics. Therefore, classical simulations based on tensor networks \cite{Banuls:2018jag, Banuls:2019rao, Dalmonte_2016,Magnifico_2021} or Markov Chain Monte Carlo face hard numerical problems \cite{robaina2020simulating, Banuls:20198n}. Quantum simulations are a promising solution, with proof-of-concept demonstrations in 1D \footnote{Here and in the following, the considered dimensions are spatial only. As such, `1D' indicates a lattice in one spatial dimension, while `2D' a lattice in two spatial dimensions. Indeed, since we use a Hamiltonian formulation of LGTs, time is not discretized.} gauge theories \cite{Mil:2019pbt, martinez2016real, muschik2017u, klco2018quantum, klco20192, bauls2019simulating, kokail2019self, schweizer2019floquet, gorg2019realization, yang2020observation, lu2019simulations} already being achieved. Extending quantum simulations of fundamental particle interactions to higher spatial dimensions represents an enormous scientific opportunity to address open questions which lie beyond the capabilities of classical computations. However, the step from one to two (or higher) spatial dimensions is extremely challenging due to an inherent increase of complexity of the models. Moreover, current limitations of so-called Noisy Intermediate-Scale Quantum (NISQ) devices \cite{preskill2018quantum} make this leap even more difficult. In this work, we develop a protocol that ease the requirements on the quantum hardware and as a consequence allows for quantum simulations of gauge theories in higher spatial dimensions.\\
%
%
%
\begin{figure*}
	\includegraphics[width=1.75\columnwidth]{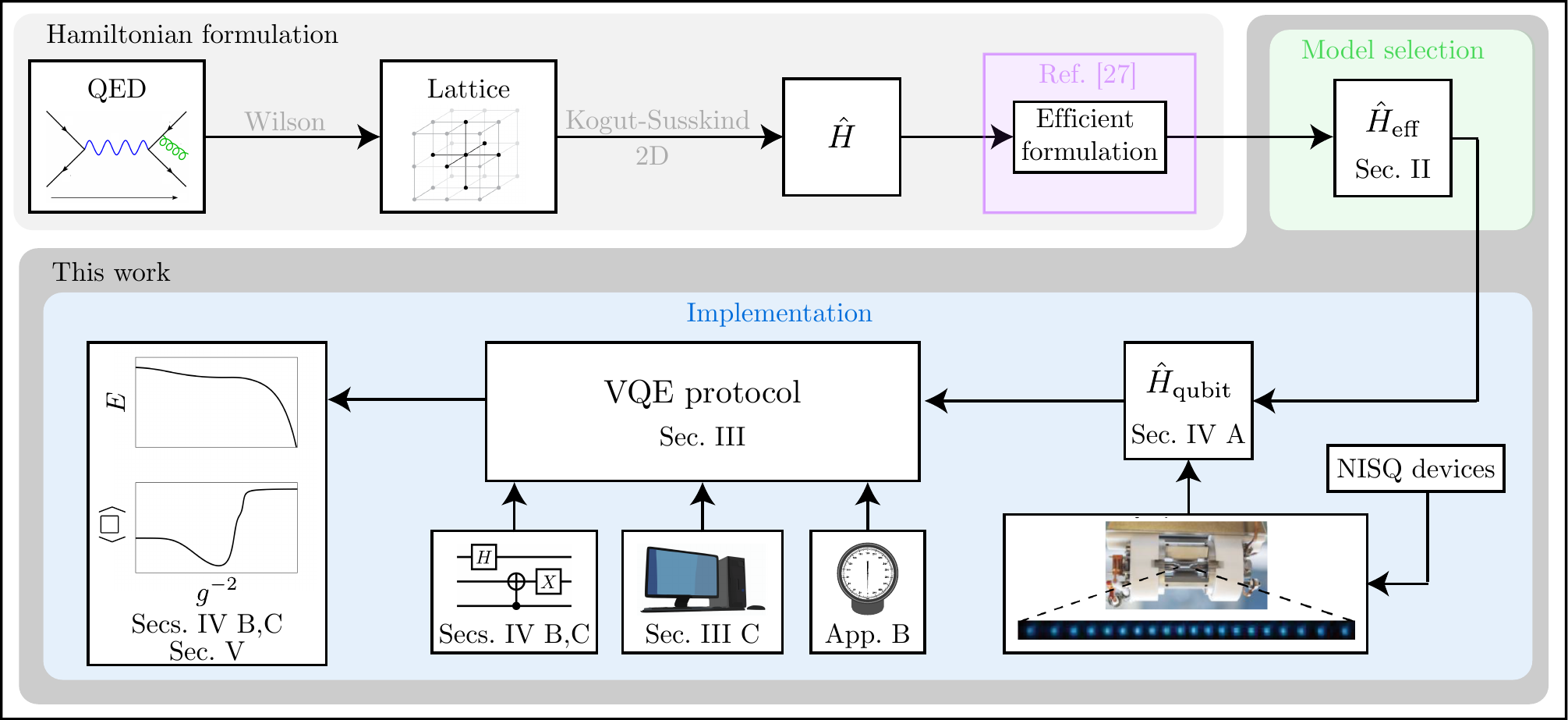}
\caption{
Blueprint for quantum simulations of lattice gauge theories. As described in Sec.~\ref{sec:main_results}, we use a VQE to simulate QED. The ingredients to perform the VQE are quantum circuits (Secs.~\ref{subsec:trappedIonResults_openBoundaryConditions} and \ref{subsec:trappedIonResults_periodicBoundaryConditions}), a classical optimizer (Sec.~\ref{sec:main_results}), and a measurement scheme to estimate the energy of a quantum state (App.~\ref{app:optimalPartitioning}). As described in Sec.~\ref{subsec:variationalQuantumSimulation_qubitEncoding}, the VQE requires a Hamiltonian $\hat{H}_{\rm qubit}$ cast in terms of qubit operators. $\hat{H}_{\rm qubit}$ is optimized for ion trap quantum computers (see picture), but can be used with any platform. In turn, $\hat{H}_{\rm qubit}$ is obtained from an effective Hamiltonian $\hat{H}_{\rm eff}$, which is derived in Sec.~\ref{sec:encodedHamiltonian} from the results of Ref.~\cite{paper1}. As schematically indicated in the ``Hamiltonian formulation'' box, Ref.~\cite{paper1} uses the Kogut-Susskind \cite{kogut1975hamiltonian} formulation of 2D LGTs whose continuum limits corresponds to QED. The outcomes of simulated VQEs are presented in Secs.~\ref{subsec:trappedIonResults_openBoundaryConditions} and \ref{subsec:trappedIonResults_periodicBoundaryConditions}, while the corresponding analytical results in Sec.~\ref{sec:quantumSimulationOf2DEffects}. Images of the lattice and ion trap quantum computer are from Refs.~\cite{lattice_img} and \cite{kokail2019self}, respectively.
}
	\label{fig:paperStructure}
\end{figure*}
\\Current quantum simulation protocols include analog, digital, and variational schemes \cite{bauls2019simulating, wiese2013ultracold,Ferguson2021}. Analog protocols \cite{Zohar_2013, kasper2016schwinger, zohar2012simulating, rico2018so, marcos2014two, hauke2013quantum, celi2020emerging, tagliacozzo2013simulation, davoudi2020towards,PhysRevD.95.094507, Kuno_2015, PhysRevLett.111.115303, PhysRevA.94.063641} aim at implementing the Hamiltonian of the simulated theory directly on quantum hardware. While this approach is challenged by the current practical difficulty of implementing gauge-invariance and the required many-body terms in the lab, first promising proof-of-concept demonstrations have been realized \cite{Mil:2019pbt, yang2020observation, schweizer2019floquet, gorg2019realization}. Noteworthy are approaches based on ultracold atoms, which allow for large system sizes and have the advantage that fermions can be used to simulate matter fields. Digital protocols \cite{bender2018digital, martinez2016real, muschik2017u, klco20192, Shaw_2020,brower2020lattice} face similar practical challenges, but have the advantage to be universally programmable and allow for the simulation of both real-time dynamics and equilibrium physics. Lastly, hybrid quantum-classical variational approaches \cite{kokail2019self, klco2018quantum, dumitrescu2018cloud, shehab2019toward} are in an early development stage and can be used to address equilibrium phenomena. These schemes do not require the simulated Hamiltonian to be realized on the quantum device. Along with their inherent robustness to imperfections \cite{mcclean2016theory}, this feature makes variational schemes suitable for NISQ technology.\\
\\As shown in Fig.~\ref{fig:paperStructure}, we use a variational approach to simulate ground state properties of lattice gauge theories (LGTs). In contrast to previous schemes, our protocols provide the novel opportunity to use existing quantum hardware \cite{kokail2019self, nam2020ground} to simulate 2D effects in LGTs, including dynamical matter and non-minimal gauge field truncations, with a perspective to go to the continuum limit. We consider quantum electrodynamics (QED), the gauge theory describing charged particles interacting through electromagnetic fields. In contrast to 1D QED, where the gauge field can be eliminated \cite{martinez2016real,muschik2017u,hamer1982massive, hamer1997series}, in higher dimensions both appear non-trivial magnetic field effects and the Fermi statistics of the matter fields become important. As a consequence, many-body terms appear in the Hamiltonian and implementations on currently available quantum hardware become challenging. In this work, we outline novel approaches to overcome these difficulties and to render near-term demonstrations possible.\\
\\Specifically, we provide effective simulation techniques for simulating quasi-2D and 2D lattice-QED systems
with open and periodic boundary conditions. To address the problem of efficiently finding the ground state of these models on NISQ
hardware, we develop a variational quantum eigensolver (VQE) algorithm \cite{mcclean2016theory} for current qubit-based quantum computers. 
In the quest of simulating LGT employing this VQE algorithm, we address the crucial points of
\begin{enumerate}
    \item developing a formulation of the problem within the resources of NISQ devices,
    \item implementing the model efficiently on the quantum hardware,
    \item having a clear procedure to scale up to larger, more complex systems, and
    \item verifying the results in known parameter regimes.
\end{enumerate}
For the first step, we cast the model into an effective Hamiltonian description, as done in Ref.~\cite{paper1}. The total Hilbert space is then reduced to a smaller gauge-invariant subspace by eliminating redundant degrees of freedom (at the cost of introducing non-local interactions). In order to measure this effective Hamiltonian on the quantum hardware (second step), we find an encoding for translating fermionic and gauge operators into qubit operators. The quantum circuits for the VQE are subsequently determined, respecting the symmetries of both the encoding and the Hamiltonian. From one side, this allows for an optimal exploration of the subsector of the Hilbert space in which gauge-invariant states lie. From the other, our circuits are Hamiltonian inspired, and can be scaled up to bigger systems and to less harsh truncations (third step). The fourth and last step in the list above is taken care by comparing the outcomes of the VQE algorithm with analytical results, in parameter regimes that are accessible to both. In the following, we resort to perturbation theory and exact diagonalization, but more advanced techniques \cite{PhysRevD.90.074501,PhysRevD.86.094504,PhysRevD.81.094505} can be in principle used.\\
\\Our work allows to run quantum simulations of 2D lattice gauge theories on lattices with arbitrary values of the lattice spacing $a$. With the results in Ref.~\cite{paper1}, it is now possible to reach a well-controlled continuum limit $a \rightarrow 0$ while avoiding the problem of autocorrelations inherent to Markov Chain Monte Carlo (MCMC) methods. This offers the exciting long-term perspective to compute physical (i.e., continuum) quantities, such as bound state spectra, non-perturbative matrix elements, and form factors that can be related to collider and low energy experiments. Studying these physical observables requires large lattices whose simulation is inaccessible to present-day quantum devices. However, various local quantities describe fundamental properties of a theory, and can be simulated on small lattices that are accessible today. An example is the plaquette expectation value, as used in the pioneering work of Creutz \cite{creutz1983monte}. Despite its simplicity, this observable can be related to the renormalized running coupling that we consider below. We emphasize that the methods presented here can be used as a launch pad to further developments in the realm of quantum simulations of LGTs, that are aiming at higher dimensions, larger lattice sizes and/or non-abelian theories. Furthermore, the LGT models can be extended to include topological terms, or finite fermionic chemical potentials, that are currently hard or even inaccessible to MCMC due to the sign problem.\\
\\We treat the case of open boundary conditions within a ladder system [see Fig.~\ref{fig:openBoundaryConventions}(a)]. Although this
system does not encompass the physics of the full 2D plane, it allows one to explore magnetic field properties
on currently available quantum hardware. For that reason, we provide a simulation protocol for the
basic building block of 2D LGTs, a single plaquette including matter, that demonstrates dynamically generated gauge fields. This is shown by observing the effect of particle-antiparticle pair creations on
the magnetic field energy. In particular, both positive and negative fermion masses are considered. In the
latter case, MCMC methods cannot be applied due to the zero mode problem \cite{Troyer:2004ge, Gattringer:2016kco}.\\
\\Additionally, we consider a single plaquette with periodic boundary conditions (see Fig.~\ref{fig:periodicBoundaryConventions}) to demonstrate
that our approach provides an important first step towards calculating the so-called ``running coupling" in gauge theories \cite{aoki2020flag, aoki2017review}. The running of the coupling, i.e., the dependence of the charge on the energy scale on which it is probed, is fundamental to gauge theories and is absent in 1D QED. For example, its precise determination in quantum chromodynamics is crucial for analyzing particle collider experiments. Here, we propose a first proof-of-concept quantum simulation of the running coupling for pure gauge QED.\\
\\
The paper is structured as follows (see Fig.~\ref{fig:paperStructure}). In Sec.~\ref{sec:encodedHamiltonian}, we introduce lattice QED and present the effective Hamiltonian description on which the proposal is based. The VQE algorithm is then outlined in Sec.~\ref{sec:main_results}, along with a short description of the available quantum hardware on which it may be implemented, along with possible classical optimization routines. In Sec.~\ref{sec:main_result}, we demonstrate our VQE algorithm for an ion-based quantum computer. The all-to-all connectivity available on this platform is an excellent match for our approach in the NISQ era, since the elimination of redundant degrees of freedom results in non-local interactions. We present a detailed experimental proposal, along with a classical simulation of the proposed experiments to demonstrate that an implementation with present-day quantum computers is feasible. Finally, we describe the physical 2D phenomena that we aim to study in Sec.~\ref{sec:quantumSimulationOf2DEffects}. Conclusions and outlooks are presented in Sec.~\ref{sec:conclusionsAndOutlook}.
\section{Simulated models}
\label{sec:encodedHamiltonian}
\noindent In this section, we present the models to be simulated in our proposal. In Sec.~\ref{subsec:encodedHamiltonian_LatticeQEDIn2+1Dimensions}, we review the Hamiltonian formulation of lattice QED in $2$ dimensions along with the truncation applied to gauge degrees of freedom. The specific systems considered in the rest of this paper are then described in Sec.~\ref{subsec:encodedHamiltonian_effectiveHamiltonianForOpenBoundaryConditions} [open boundary conditions (OBC)] and Sec.~\ref{subsec:encodedHamiltonian_effectiveHamiltonianForPeriodicBoundaryConditions} [periodic boundary conditions (PBC)].
\subsection{Lattice QED in 2 dimensions}
\label{subsec:encodedHamiltonian_LatticeQEDIn2+1Dimensions}
\begin{figure}[t]
  \includegraphics[width=\columnwidth]{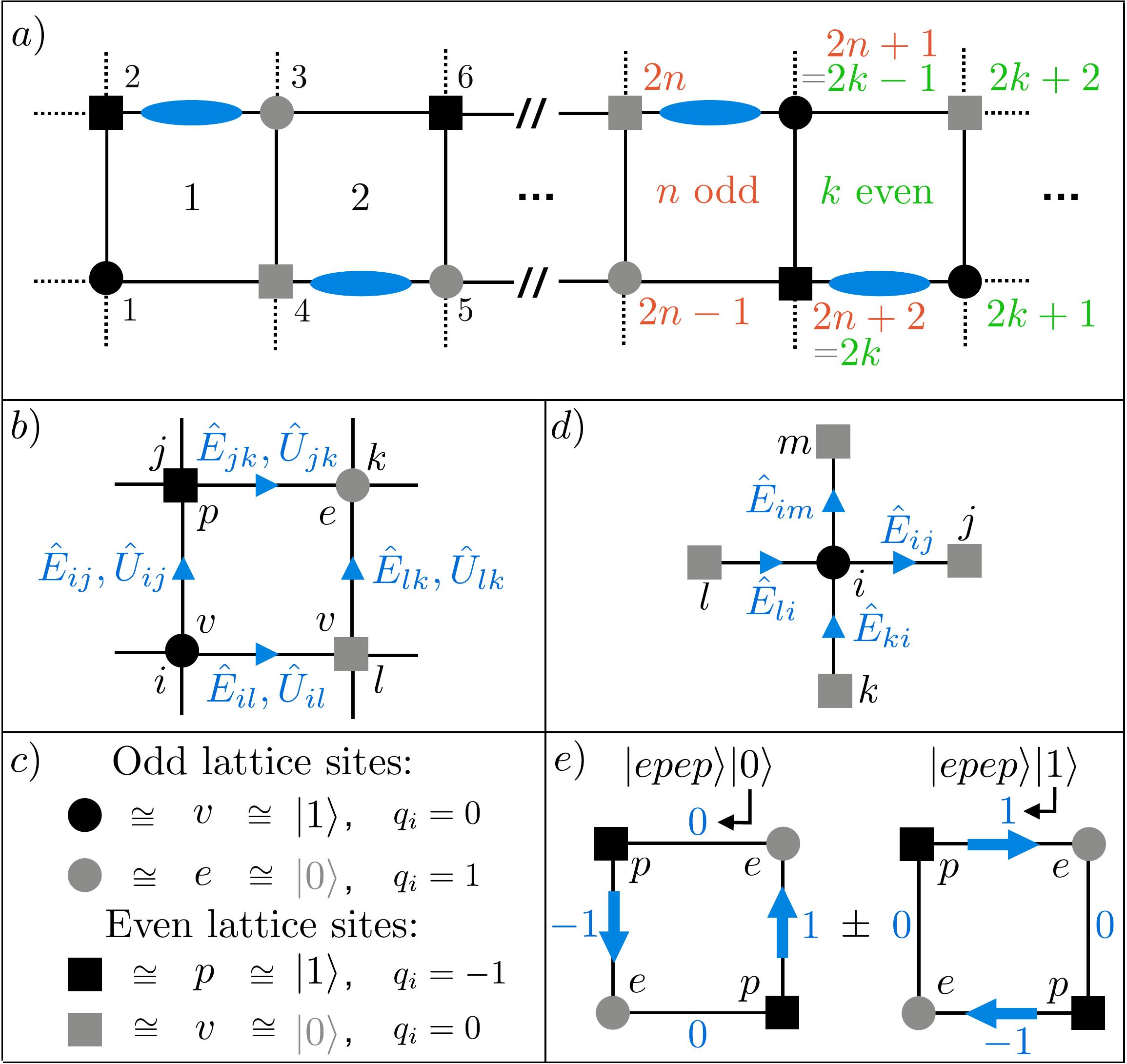}
  \caption{Conventions for lattice QED in 2D. \textbf{(a)} Ladder system with open boundary conditions. Using Gauss' law, the number of gauge degrees of freedom can be reduced to one per plaquette (blue ellipses). \textbf{(b)} A single plaquette with matter sites at the vertices and gauge fields on the links. Circles (squares) represent odd (even) sites, and black (grey) corresponds to unoccupied (occupied) fermionic fields. Positive field direction is to the right and up. \textbf{(c)} Table showing the mapping between fermionic sites, particles ($e$) and antiparticles ($p$), and spins. \textbf{(d)} Conventions for Gauss' law. \textbf{(e)} The two gauge-field configurations that minimize the electric field energy for a single plaquette with two particle-antiparticle pairs. As explained in the main text, these configurations are relevant for the 2D effects discussed in Sec.~\ref{subsec:quantumSimulation_openBoundaryConditions} (see App.~\ref{app:newAppendix} for more details).}
  \label{fig:openBoundaryConventions}
\end{figure}
\noindent In this work, we consider two-dimensional lattices, with matter and gauge fields defined on the vertices and on the links, respectively. Using staggered fermions \cite{kogut1975hamiltonian}, electrons and positrons are represented by single component fermionic field operators $\hat{\phi}_i$ for each site $i$. As shown in Figs.~\ref{fig:openBoundaryConventions}(b) and \ref{fig:openBoundaryConventions}(c), odd (even)-numbered lattice sites hold particles (antiparticles), that carry a $+1$ $(-1)$ charge $q_i$.
%
Gauge fields on the links between sites $i$ and $j$ are described by the operators $\hat{E}_{ij}$ (electric fields) and $\hat{U}_{ij}$. Electric field operators take integer eigenvalues $e_{ij} = 0,~\pm 1,~\pm 2, \dots$ with $\hat{E}_{ij}\ket{e_{ij}} = e_{ij}\ket{e_{ij}}$, while $\hat{U}_{ij}$ acts as a lowering operator on electric field eigenstates, i.e., $\hat{U}_{ij} \ket{e_{ij}} = \ket{e_{ij} -1}$, with $[\hat{U}_{ij}, \hat{U}^{\dag}_{kl}] = 0$ and $[\hat{E}_{ij}, \hat{U}_{kl}] = -\delta_{i,k} \delta_{j,l}\hat{U}_{ij}$.\\
\\Using the Kogut-Susskind formulation \cite{kogut1975hamiltonian}, the Hamiltonian consists of an electric, a magnetic, a mass, and a kinetic term; $\hat{H} = \hat{H}_{\textrm{E}}  + \hat{H}_{\textrm{B}} + \hat{H}_{\textrm{m}} + \hat{H}_{\textrm{kin}}$, where
\begin{subequations}\label{eq:HtotOpenBoundary}
  \begin{align}
    \hat{H}_{\textrm{E}} =& \frac{g^2}{2} \sum_{ \substack{i, \\ i \stackrel{+}\longrightarrow j}} \hat{E}^2_{ij}, \label{eq:HEOpenBoundary}\\
    \hat{H}_{\textrm{B}} =& -\frac{1}{2g^2a^2} \sum_{n = 1}^{N}\Big(\hat{P}_n + \hat{P}^{\dag}_n\Big), \label{eq:HBOpenBoundary} \\
    \hat{H}_{\textrm{m}} =& m \sum_{i \in \textrm{sites}} (-1)^{i+1}\hat{\phi}_i^{\dag} \hat{\phi}_i, \label{eq:HmOpenBoundary} \\
    \hat{H}_{\textrm{kin}} =& \Omega\Big( \sum_{\substack{i~\textrm{odd},\\i \stackrel{+}\longrightarrow j}} \hat{\phi}_i^{\dag} \hat{U}_{ij} \hat{\phi}_j \nonumber \\&+ \sum_{\substack{i~\textrm{even}, \\i \stackrel{+}\longrightarrow j}} \hat{\phi}_i \hat{U}^{\dag}_{ij} \hat{\phi}_j^{\dag} \Big) + \textrm{H.c.}. \label{eq:HkinOpenBoundary}
  \end{align}
\end{subequations}
In the summations, we use $i \stackrel{+} \longrightarrow j$ to denote the link between lattice sites $i$ and $j$ with positive orientation [see Fig.~\ref{fig:openBoundaryConventions}(a) and (b)]. We denote the bare coupling by $g$, $m$ is the fermion mass, $a$ the lattice spacing and $\Omega $ the kinetic strength. 
We use natural units $\hbar = c = 1$ and all operators in Eqs.~\eqref{eq:HtotOpenBoundary} are dimensionless \cite{paper1}.\\
%
\\In the Hamiltonian above, we introduced the operator $\hat{P}_n = \hat{U}_{ij}^{\dag} \hat{U}_{jk}^{\dag} \hat{U}_{il} \hat{U}_{lk}$, where sites $(i,j,k,l)$ form a closed loop clockwise around the plaquette $n$ as in Fig.~\ref{fig:openBoundaryConventions}(b) \cite{wilson1974confinement}. This allows us to define the plaquette operator
\begin{align} \label{eq:PlaquetteOperator}
  \Box &= \frac{1}{2N} \sum_{n = 1}^N\Big(\hat{P}_n + \hat{P}_n^{\dag}\Big),
\end{align}
with $N$ being the number of plaquettes.\\
\\At each vertex $i$, gauge invariance is imposed by the symmetry generators \cite{kogut1975hamiltonian, Zohar_2013} $\hat{G}_i = \hat{E}_{li} - \hat{E}_{ij} + \hat{E}_{ki} - \hat{E}_{im} - \hat{q}_i$ [see Fig.~\ref{fig:openBoundaryConventions}(d) for the definition of $l$, $j$, $k$, $m$], where $\hat{q}_i = \hat{\phi}_{i}^{\dag} \hat{\phi}_{i} - \frac{1}{2}\left[1 + (-1)^{i}\right]$ is the charge operator. Relevant, i.e., gauge-invariant quantum states are defined by Gauss' law $\hat{G}_{i}\ket{\Psi_{\textrm{phys}}} = \epsilon_{i}\ket{\Psi_{\textrm{phys}}}$ for each vertex of the lattice, where the eigenvalue $\epsilon_{i}$ corresponds to the static charge at site $i$. We consider the case $\epsilon_i = 0~\forall i$, but background charges can be easily included in the derivations below. In the continuum limit $a \rightarrow 0$ and three spatial dimensions, Gauss' law takes the familiar form $\nabla \hat{E} = \hat{\rho}$, where $\hat{\rho}$ is the charge density.\\
\\Gauss' law can be used to lower the requirements for a quantum simulation.
More specifically, within the Hamiltonian formulation of a gauge theory, only an exponentially small part of the whole Hilbert space consists of gauge-invariant states. In QED, this subspace is selected by applying Gauss' law with a specific choice of the static charges.
As a result, it is possible to eliminate some of the gauge fields and obtain an effective Hamiltonian which is constrained to the chosen subspace (see also Ref.~\cite{bender2020gauge}).
Practically, this reduces the number of qubits that are required to simulate the gauge fields -- which is particularly important in the NISQ era.
As can be seen in the following sections, the trade-off of this resource-efficient encoding is the appearance of long-range many-body interactions in the effective Hamiltonian. 
Nevertheless, universal quantum computers offer the possibility to perform gates between arbitrary pairs of qubits (long-range interactions). Moreover, the VQE approach is promising when dealing with interactions that cannot be directly implemented in the quantum hardware. Therefore, 
resorting to the effective Hamiltonians of the models considered for our quantum simulation in Sec.~\ref{sec:main_result} is advantageous in terms of experimental feasibility of our protocols.
\subsection{Effective Hamiltonian for open boundary conditions}
\label{subsec:encodedHamiltonian_effectiveHamiltonianForOpenBoundaryConditions}
\noindent In this section, we consider a one-dimensional ladder of plaquettes with OBC [see Fig.~\ref{fig:openBoundaryConventions}(a)]. Although this system does not encapsulate the full 2D physics of QED, it allows us to study important aspects of gauge theories that are not present in one spatial dimension, such as magnetic phenomena.\\
\\The OBC for a ladder of $N$ plaquettes appear as dashed lines in Fig.~\ref{fig:openBoundaryConventions}(a), indicating null background field. While there are $3N+1$ gauge fields in the ladder, applying Gauss' law to all vertices reduces the independent gauge degrees of freedom to $N$. Given the freedom in choosing the independent gauge fields, we select the $(2n, 2n+1)$ links for each plaquette $n$, as shown by the blue ellipses in Fig.~\ref{fig:openBoundaryConventions}(a). On each link that does not hold an independent gauge field, the corresponding unitary operator $\hat{U}$ is set to the identity.
As a result, the plaquette operator becomes $\hat{P}_n = \hat{U}_{2n, 2n+1}$ and both the kinetic and magnetic terms are consequently simplified. The effective Hamiltonian for the ladder is derived in App.~\ref{app:ElectricEnergyContribution}.\\
%
\\We now focus on the basic building block of the 2D ladder system, the plaquette [see Fig.~\ref{fig:openBoundaryConventions}(b)]. Setting $N=1$ in Eqs.~\eqref{eq:OpenBoundarySpinHamiltonian}, the effective Hamiltonian for a single plaquette with OBC becomes
\begin{subequations} \label{eq:singlePlaquetteHamiltonian}
  \begin{align}
    \hat{H}_{\textrm{E}} = &\frac{g^2}{2}\Big[\Big(\hat{E}_{23}\Big)^2 + \Big(\hat{E}_{23} + \hat{q}_2\Big)^2 \nonumber \\
    &+ \Big(\hat{E}_{23} - \hat{q}_{3}\Big)^2 + \Big(\hat{E}_{23} + \hat{q}_{1} + \hat{q}_{2}\Big)^2 \Big],  \label{eq:singlePlaquetteHamiltonianHE}\\
    \hat{H}_{\textrm{B}} = &-\frac{1}{2g^2}\Big(\hat{U}_{23} + \hat{U}^\dagger_{23}\Big), \\
    \hat{H}_{\textrm{m}} =& m \sum_{i =1}^4 (-1)^{i+1}\hat{\phi}_i^{\dag} \hat{\phi}_i, \label{eq:singlePlaquetteHamiltonianHm}\\
    \hat{H}_{\textrm{kin}} = &\Omega \Big(\hat{\phi}_1^{\dag} \hat{\phi}_2 + \hat{\phi}_1^{\dag} \hat{\phi}_4 +  \hat{\phi}_2 \hat{U}^\dagger_{23} \hat{\phi}_3^{\dag} + \hat{\phi}_4 \hat{\phi}_3^{\dag}\Big) + \textrm{H.c.}.   \label{eq:singlePlaquetteHamiltonianHkin}
  \end{align}
\end{subequations}
The notation for describing the state of the plaquette with OBC is a tensor product of two kets, with the first ket representing the matter sites $1-4$ ($v$, $e$, and $p$ are the vacuum, particle, and antiparticle, respectively) and the second ket the state of the gauge field on the $(2,3)$ link [see Fig.~\ref{fig:openBoundaryConventions}(e) for examples].\\
\\We remark that in 1D QED, all gauge fields can be eliminated for systems with OBC. As a result, only two-body terms remain in the Hamiltonian \cite{hamer1982massive, hamer1997series, martinez2016real, muschik2017u}. For the plaquette, many-body terms are unavoidable since gauge degrees of freedom survive. This represents one of the main challenges in simulating LGTs in more than one spatial dimensions.
\subsection{Effective Hamiltonian for periodic boundary conditions}
\label{subsec:encodedHamiltonian_effectiveHamiltonianForPeriodicBoundaryConditions}
\begin{figure}[t]
  \centering
  \includegraphics[width=\columnwidth]{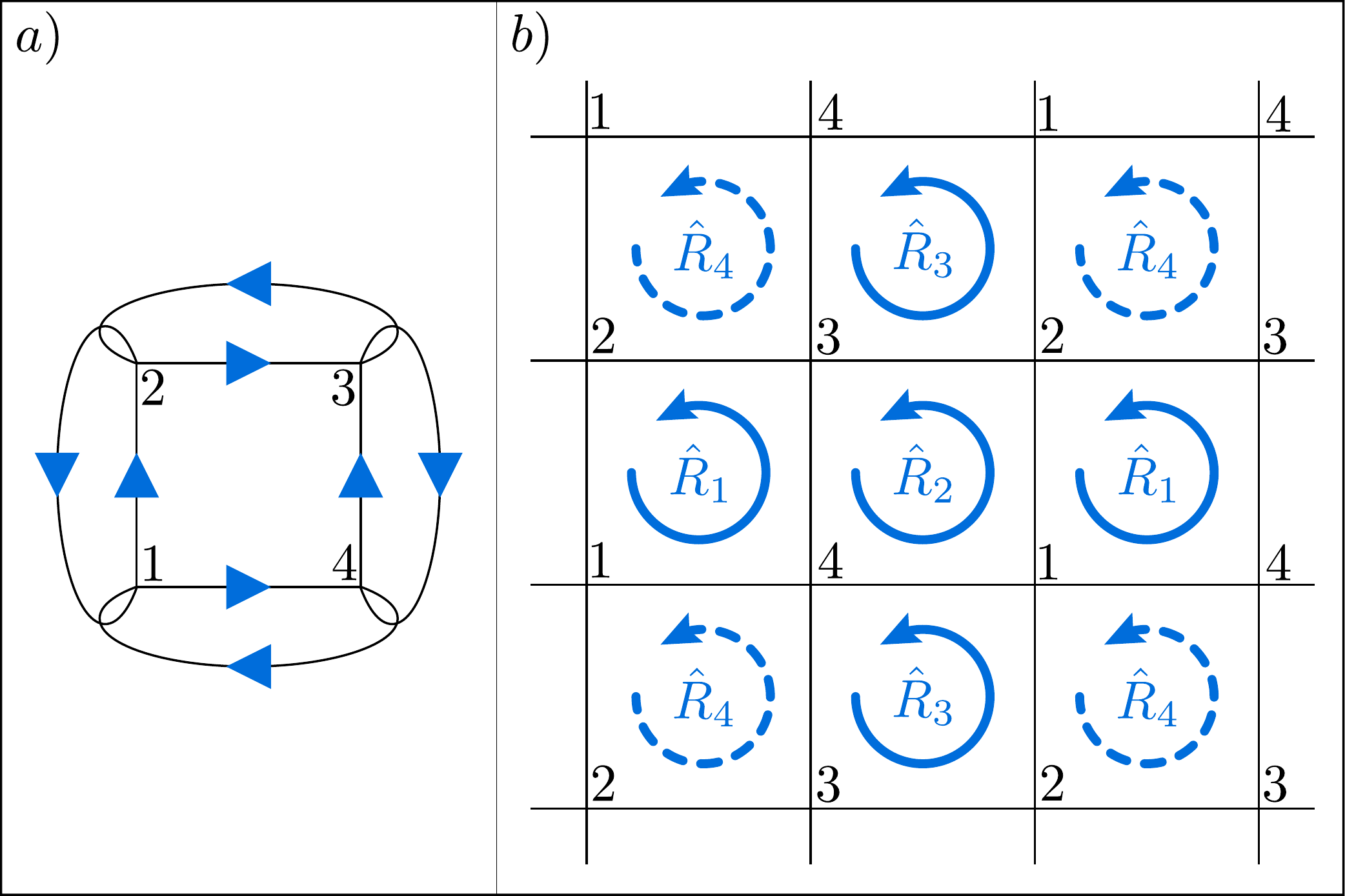}
  \caption{Single periodic boundary plaquette for a pure gauge theory. \textbf{(a)} Representation in terms of the eight gauge fields (blue arrows), that are associated with the links of the lattice. \textbf{(b)} Representation in terms of independent gauge-invariant (i.e., physical) operators called ``rotators''. As explained in Ref.~\cite{paper1}, the plaquette can be seen as an infinite lattice of plaquettes. Moreover, to explore ground state properties, only three independent rotators $\hat{R}_1$, $\hat{R}_2$, and $\hat{R}_3$ (shown in solid blue) are sufficient for describing the system.}
  \label{fig:periodicBoundaryConventions}
\end{figure}
\noindent For the second model, we consider a square lattice with PBC and without fermionic matter. This system has been considered in Ref.~\cite{paper1}, but for the benefit of the reader we summarize the main results that are required for the VQE simulation. As explained in Sec.~\ref{subsec:encodedHamiltonian_LatticeQEDIn2+1Dimensions}, Gauss' law can be used to eliminate redundant gauge fields, resulting in an unconstrained effective Hamiltonian. 
%
The basic building block, a single plaquette with PBC, includes eight gauge fields [see Fig.~\ref{fig:periodicBoundaryConventions}(a)], and is equivalent to an infinite 2D lattice of four distinct plaquettes [see Fig.~\ref{fig:periodicBoundaryConventions}(b)]. 
Due to the absence of matter, we consider the pure gauge Hamiltonian $\hat{H}_{\textrm{gauge}} = \hat{H}_{\textrm{E}} + \hat{H}_{\textrm{B}}$, given by Eqs.~\eqref{eq:HEOpenBoundary} and \eqref{eq:HBOpenBoundary}.\\
\\Since we are interested in studying ground state properties, it is sufficient to consider three independent gauge degrees of freedom (see Ref.~\cite{paper1}), called ``rotators" $\hat{R}_{i}$, $i \in \{1,2,3\}$ shown in Fig.~\ref{fig:periodicBoundaryConventions}(b). Rotators are a convenient representation for the gauge fields, as they correspond to the circulating electric field in a specific plaquette. As such, the operators $\hat{P}_{n}$ ($\hat{P}^{\dag}_{n}$) in Eqs.~\eqref{eq:HBOpenBoundary} and \eqref{eq:PlaquetteOperator} are directly their descending (ascending) operators. Indeed, the commutation relations between a rotator $\hat{R}$ and its associated operator $\hat{P}$ are the same as the ones of an electric field $\hat{E}$ and its lowering operator $\hat{U}$, presented in Sec.~\ref{subsec:encodedHamiltonian_LatticeQEDIn2+1Dimensions}. The notation for describing the state of the plaquette with PBC is thus a tensor product of three kets, each corresponding to one of the three rotators represented by full lines in Fig.~\ref{fig:periodicBoundaryConventions}(b).\\
\\Usually, LGT Hamiltonians are formulated in the basis of the electric field, where $\hat{H}_{\textrm{E}}$ in $\hat{H}_{\textrm{gauge}}$ is diagonal. For large values of the bare coupling $g$, the term $\hat{H}_{\textrm{E}}$ is dominant, and this representation is efficient. 
For small $g$, however, the magnetic term $\hat{H}_{\textrm{B}}$ is dominant, and the eigenstates of $\hat{H}_{\textrm{gauge}}$ are superpositions of all electric field basis states. 
Since the Hilbert space of these operators is of infinite dimension, this necessarily leads to truncation errors when the Hamiltonian is mapped to any quantum device (as in Sec.~\ref{subsec:variationalQuantumSimulation_qubitEncoding}).
To mitigate the truncation error, we apply the novel technique introduced in Ref.~\cite{paper1}, and resort to two different formulations of the Hamiltonian. The electric formulation in Eqs.~\eqref{eq:HtotOpenBoundary} is applied in the region $g^{-2} \lesssim 1$, while for $g^{-2} \gtrsim 1$ a so-called magnetic basis is used, for which $\hat{H}_{\textrm{B}}$ is diagonal (see also Ref.~\cite{stryker2018gauss}). 
This magnetic representation is obtained using the cyclic group $\mathbb{Z}_{2L+1}$ ($L \in \mathbb{N}$) to approximate the continuous group U(1) associated to each rotator. We first take an approximation in the form of a series expansion that is exact for $L \rightarrow \infty$. Then, we apply the Fourier transform and truncate the group $\mathbb{Z}_{2L+1}$ by only considering $2l+1 < 2L+1$ states in order to achieve equivalence to a truncated U(1) theory. As such, the rotator's basis becomes $\ket{-l},...,\ket{0},...,\ket{l}$. This leaves one with $L$ as a free parameters, whose optimal choice, along with the technical details of approximating the U(1) theory with a truncated $\mathbb{Z}_{2L+1}$ model, is studied in Ref.~\cite{paper1}.\\
%
\\We thus arrive at the Hamiltonian $\hat{H}_{\textrm{gauge}}^{(\gamma)} = \hat{H}_{\textrm{E}}^{(\gamma)} + \hat{H}_{\textrm{B}}^{(\gamma)}$, where $\gamma = e$ indicates the electric representation, 
\begin{subequations} \label{eq:periodicHamiltonianElectric}
  \begin{align}
    \hat{H}_{\textrm{E}}^{(e)} = &~2g^2 \Big[(\hat{R}_1)^2 + (\hat{R}_2)^2 + (\hat{R}_3)^2 \nonumber\\&- \hat{R}_2 (\hat{R}_1 + \hat{R}_3)\Big], \label{eq:periodicHamiltonianElectricE}\\
     \hat{H}_{\textrm{B}}^{(e)} = &-\frac{1}{2g^2}\Big(\hat{P}_1 + \hat{P}_2 + \hat{P}_3 \nonumber\\&+ \hat{P}_1 \hat{P}_2 \hat{P}_3 + \textrm{H.c.}\Big) \label{eq:periodicHamiltonianElectricB},
  \end{align}
\end{subequations}
and we denote the magnetic representation with $\gamma = b$,
\begin{subequations} \label{eq:periodicHamiltonianMagnetic}
  \begin{align}
    \hat{H}_{\textrm{E}}^{(b)} = &~g^2 \sum_{\nu = 1}^{2L} \Big\{f_{\nu}^c \sum_{i=1}^{3} (\hat{P}_{i})^{\nu} + \frac{f_{\nu}^{s}}{2}\big[(\hat{P}_{2})^{\nu} - (\hat{P}_{2}^\dagger)^{\nu}\big] \nonumber\\
		&~\times \sum_{\mu = 1}^{2L} f_{\mu}^{s} \big[(\hat{P}_{1})^{\mu} + (\hat{P}_{3})^{\mu}\big]\Big\} + \textrm{H.c.},\label{eq:periodicHamiltonianMagneticE}\\
    \hat{H}_{\textrm{B}}^{(b)} = &-\frac{1}{g^2} \Big[ \sum_{i = 1}^{3} \cos \Big(\frac{2 \pi \hat{R}_{i}}{2L+1}\Big)  \nonumber \\
    & + \cos \Big(\frac{2 \pi (\hat{R}_{1} + \hat{R}_{2} + \hat{R}_{3})}{2L+1}\Big)\Big]. \label{eq:periodicHamiltonianMagneticB}
  \end{align}
\end{subequations}
%
The coefficients $f_k^c$ and $f_k^s$ come from the Fourier transform, and their analytic form is
\begin{align}
	f_{\nu}^{s} = &\frac{(-1)^{\nu+1}}{2 \pi} \Big[\psi_{0} \Big(\frac{2L+1+\nu}{2(2L+1)}\Big) \nonumber \\
	& -\psi_{0} \Big(\frac{\nu}{2(2L+1)}\Big)\Big], \\
	f_{\nu}^{c} = &\frac{(-1)^{\nu}}{4 \pi^{2}} \Big[ \psi_{1} \Big(\frac{\nu}{2(2L+1)}\Big) \nonumber \\
	&- \psi_{1} \Big(\frac{2L+1+\nu}{2(2L+1)}\Big)\Big],
\end{align}
where $\psi_{k}(\cdot)$ is the $k$-th polygamma function \cite{paper1}. We note that making use of electric and magnetic representations in the case of the plaquette with OBC (see Sec.~\ref{subsec:encodedHamiltonian_effectiveHamiltonianForOpenBoundaryConditions}) is straightforward. For clarity, however, we use this formulation only when considering PBC.
\section{Foundations for variational quantum simulations}
\label{sec:main_results}
\noindent As detailed in Sec.~\ref{sec:encodedHamiltonian}, the Hamiltonians associated with the considered models contain exotic long-range interactions and many-body terms that are beyond the capabilities of current analogue quantum simulations. 
This is particularly true once the infinite dimensional operators $\hat{E}$ and $\hat{U}$, corresponding to the gauge field and its connection \cite{kogut1975hamiltonian}, respectively, are encoded in terms of qubit operators (see Sec.~\ref{subsec:variationalQuantumSimulation_qubitEncoding}). 
Resorting to digital approaches, trotterized adiabatic state preparation is possible in theory \cite{jordan2012quantum}. However, the high complexity of LGTs beyond one dimension renders this approach infeasible for the technology available today.
Due to the absence of large-scale and universal fault-tolerant quantum computers we thus employ a hybrid quantum-classical strategy, using a VQE protocol that renders the observation of the effects described in Sec.~\ref{sec:quantumSimulationOf2DEffects} within reach of present-day quantum devices. 
In fact, those protocols have already proven their validity in quantum chemistry \cite{mcclean2016theory, o2016scalable, hempel2018quantum, peruzzo2014variational}, nuclear physics \cite{dumitrescu2018cloud, shehab2019toward} and even classical applications \cite{farhi2014quantum,borle2020quantum,bravo2020variational}.\\
\\In the following, we first introduce the VQE protocol (Sec.~\ref{subsec:variationalQuantumSimulation_variationalQuantumSimulation}) and discuss feasibility on currently available quantum platforms (Sec.~\ref{subsec:trappedIonResults_trappedIons}). Then, we describe our minimization algorithm, which is run on the classical device to find the desired state (Sec.~\ref{subsec:optimizationRoutine}).
\subsection{Hybrid quantum-classical simulations}
\label{subsec:variationalQuantumSimulation_variationalQuantumSimulation}
\noindent The VQE is a closed feedback loop between a quantum device and an optimization algorithm performed by a classical computer \cite{peruzzo2014variational, mcclean2016theory}. 
The quantum device is queried for a cost function $\mathcal{C}(\boldsymbol{\theta})$, which is expressed in terms of a parameterized variational state $\ket{\Psi(\boldsymbol{\theta})} = U(\boldsymbol{\theta})\ket{\Psi_{\textrm{in}}}$. 
Here, $\ket{\Psi_{\textrm{in}}}$ is an initial state that can be easily prepared, while $U(\boldsymbol{\theta})$ entails the application of a quantum circuit involving the variational parameters $\boldsymbol{\theta}$. The classical optimization algorithm minimizes $\mathcal{C}(\boldsymbol{\theta})$, and provides an updated set of variational parameters $\boldsymbol{\theta}$ after each iteration of the feedback loop.
Since our quantum simulations aim at preparing the ground state (see Sec.~\ref{sec:quantumSimulationOf2DEffects}), our VQE cost function is defined as the expectation value of the system's Hamiltonian $\hat{H}$ with respect to the variational state, i.e. $\mathcal{C}(\boldsymbol{\theta}) = \bra{\Psi(\boldsymbol{\theta})} \hat{H} \ket{\Psi(\boldsymbol{\theta})}$.\\
\\Importantly, the Hamiltonian is never physically realized on the quantum hardware. Instead, its expectation value is estimated by measuring $\hat{H}$ with respect to the variational state $\ket{\Psi(\boldsymbol{\theta})}$ \cite{peruzzo2014variational, mcclean2016theory} (see App.~\ref{app:optimalPartitioning}). Thus, the VQE is advantageous whenever the studied model contains complicated long-range, many-body interactions (as in this case), that cannot be realized in current devices. Moreover, this approach is insensitive to several systematic errors, such as offsets in the variational parameters $\boldsymbol{\theta}$, since these are compensated by the classical optimization routine.
Both these properties relax the quantum resource requirements and facilitate future experimental implementations of the VQE routines presented in the remainder of this paper.\\

\subsection{Quantum hardware considerations}
\label{subsec:trappedIonResults_trappedIons}
\noindent 
In this work, we consider qubit-based platforms for the realization of a proof-of-principle experiment with present-day technology. Promising quantum computing platforms include configurable Rydberg arrays \cite{bernien2017probing, labuhn2016tunable}, superconducting architectures \cite{arute2019quantum, corcoles2015demonstration}, and trapped ions \cite{lanyon2011universal, debnath2016demonstration}. Comparing these approaches, Rydberg-based systems offer the advantage that $2$D and $3$D arrays can be realized using optical tweezers, which translates to a large number of available qubits. The ability to implement large qubit registers will make future generations of Rydberg arrays a very promising candidate for our schemes, once higher levels of controllability and gate fidelity become available. Superconducting and ion-based architectures offer both high controllability and gate fidelities already today. For superconducting qubits, entangling gates are inherently of nearest-nearest neighbour type, which implies that entangling operations between non-neighbouring qubits have to be realized through a number of swap gates. While this is entirely possible, especially for next-generation devices, our models entail long-range interactions that result from the elimination of redundant gauge fields (see Secs.~\ref{subsec:encodedHamiltonian_effectiveHamiltonianForOpenBoundaryConditions} and \ref{subsec:encodedHamiltonian_effectiveHamiltonianForPeriodicBoundaryConditions}) and thus suggest the necessity of creating highly entangled states which would require significant gate overhead on superconducting platforms.
In contrast, ion-based quantum computers \cite{benhelm2008towards, brown2011single} have all-to-all connectivity, allowing for addressed entangling gates between arbitrary qubits. This aligns well with our target models, motivating their use for this proposal. Despite their comparatively slow readout limits the measurement budget, we solve this issue with both an efficient measurement strategy (see App.~\ref{app:optimalPartitioning}) and an optimized classical algorithm (see Sec.~\ref{subsec:optimizationRoutine}). A realistic budget for currently available ion-based quantum computers is given by about $10^7$ measurement shots. This includes the time for the quantum computer to initialize, modify and measure the states, and the time for the classical computer to minimize the cost function $\mathcal{C}(\boldsymbol{\theta})$.\\
\\Here, we describe the properties of ion-based computers in more detail. We consider a string of ions confined in a macroscopic linear Paul trap \cite{lanyon2011universal, debnath2016demonstration}. The qubit states $\ket{0}$ and $\ket{1}$ are encoded in the electronic states of a single ion and can be manipulated using laser light in the visible spectrum. A universal set of quantum operations is available, which can be combined to implement arbitrary unitary operations. More specifically, the available gate-set consists of local rotations $\hat{U}_{j}(\phi) = \exp (-i \frac{\phi}{2} \hat{\sigma}_{j}^{\alpha})$ and addressed M\o lmer-S\o rensen (MS) gates \cite{sorensen1999quantum} between arbitrary pairs of qubits. These gates are implemented with fidelities exceeding $98\%$ for both single- and multi-qubit gates \cite{benhelm2008towards, brown2011single, zhang2017observation}.
%
%
While currently available ion-based hardware provides a sufficient number of qubits to carry our proposed protocols, future LGT quantum simulations addressing larger lattices will require large-scale quantum devices. Ion based quantum computers can be scaled up using segmented 2D traps \cite{kielpinski2002architecture} and networking approaches that connect several traps together \cite{monroe2014large, Ragg_2019} and therefore offer a pathway for developing quantum simulations of increasing size and complexity.
\subsection{Classical Optimization Routine}
\label{subsec:optimizationRoutine}
\noindent The classical optimization routine employed for the VQE needs to be chosen depending on the requirements from both the hardware and the cost function $\mathcal{C}(\boldsymbol{\theta})$. Indeed, the stochastic nature of the latter has to be taken into account, and the experimental repetition rate poses limitations on the number of data points that the classical machine can use for minimization. In the following, we consider different optimization routines, discuss their strengths and weaknesses, and motivate our specific choice.\\
\\Algorithms based on the scheme of gradient descent are promising candidates, particularly when the number of variational parameters grows large. Recently, a variety of these algorithms have been modified to take the stochastic nature of the cost function into account \cite{Kingma2014aa, Ruder2017An-overview}, and there are versions which are tailored to the quantum regime \cite{Stokes2020Quantum}. 
Methods to measure the gradient of an operator \cite{Li2017} have been proposed, but they are costly when applied to multi qubit gates and require additional ancilla qubits \cite{Schuld:2019aa}. In addition, finite step size approximations of the gradient require a large number of experimental evaluations to obtain reliable values for the gradient.\\
\\A limitation of the available quantum hardware is the small number of experimental shots, particularly when resorting to ion-based platforms.
This drawback is especially limiting for gradient based optimization. Indeed, to prevent the VQE to settle in a local minimum, one has to perform multiple runs from different initial conditions. Furthermore, already acquired data cannot easily be recycled in scenarios where the algorithm is applied to different parameters regimes (e.g. variations in $g$ in Sec.~\ref{sec:main_result}). A family of algorithms which can recycle previous data is the mesh-based. These evaluate the cost function on a grid in parameter space, with refinements where the global minimum is expected to be \cite{Audet2006Mesh}. Mesh and values of the cost function can be stored, and reused for a different set of parameters, to accelerate the optimization.
The feature of building a data repository is shared with Bayesian optimizers, which build a regression model of the energy landscape based on Gaussian processes \cite{Rasmussen2003}. These algorithms explore the variational parameters' space by performing cost function evaluations at the location of potential minimizing points, suggested by their regression model. Hence, they are able to reduce the required number of measurements substantially. However their effectiveness is limited to around $20$ variational parameters \cite{Frazier2018aa}.\\
\\Here, we employ an optimization algorithm similar to the one used in Ref.~\cite{kokail2019self}, which is a modified version of the Dividing Rectangles algorithm (DIRECT) \cite{hooke1961direct,JonesOriginalDirect93, DirectConvergence2004, NicholasDirect2014, LiuDirect2015}. This algorithm divides the search space into so-called hypercells. Each hypercell contains a single sample point representative of the cost function value in that cell. The algorithm selects promising hypercells, to be divided into smaller cells, based on the cost function value as well as the cell size. Larger cells contain more unexplored territory and are hence statistically more likely to harbour the global minimum. During the optimization, the algorithm maintains a regression model as used in Bayesian optimization. This metamodel is used for an accurate function value estimation that aids in selecting the hypercells to be divided. Furthermore, at regular intervals, one or more direct Bayesian optimization steps are carried out \cite{Frazier2018aa}. 
Our choice is motivated from the small number of measurement shots that can be performed on an ion-based quantum computer (see Sec.~\ref{subsec:variationalQuantumSimulation_variationalQuantumSimulation}), and from the limited number of variational parameters $\boldsymbol{\theta}$ required in our models (see Sec.~\ref{sec:main_result}).
\section{Quantum simulation of 2D LGTs}\label{sec:main_result}
\noindent In this section, we present the VQE protocol for simulating the two models introduced in Sec.~\ref{sec:encodedHamiltonian}. We give numerical results from a classical simulation, which includes the projection noise error. 
Both proposed experiments prepare the ground state of the theory and measure the ground state expectation value of the plaquette operator $\expval{\Box} \sim \expval{\hat{H}_\mathrm{B}}$, as defined in Sec.~\ref{subsec:encodedHamiltonian_LatticeQEDIn2+1Dimensions}. 
For OBC, this allows to study the dynamical generation of gauge fields by pair creation processes, while in the case of PBC it is related to the renormalization of the coupling at different energy scales (see Sec.~\ref{sec:quantumSimulationOf2DEffects}).
\subsection{Encoding for quantum Hardware}
\label{subsec:variationalQuantumSimulation_qubitEncoding}
\noindent To run a VQE, we first require an encoding of the models outlined in Sec.~\ref{sec:encodedHamiltonian} for qubit-based quantum hardware. 
While we follow the Jordan-Wigner transformation \cite{jordan1928pauli} for mapping the fermionic operators, there is not a unique procedure to represent the gauge field operators (alternatives are given in Refs. \cite{lewis2019qubit, Shaw_2020}). Here, the encoding is chosen to reduce the complexity of the quantum circuits and to respect symmetries that both suit the simulated models and the quantum platform.\\
\\Gauge field operators are defined on infinite dimensional Hilbert spaces and the gauge field takes the values $0,\,\pm1,\,\pm2, \dots$ (see Sec.~\ref{subsec:encodedHamiltonian_LatticeQEDIn2+1Dimensions}).
To simulate them using finite-dimensional quantum systems, a truncation scheme is required.
Let us take $2l+1$ basis states into account, i.e. the gauge field can take at most the values $\pm l$. Consequently, we substitute the electric field operator $\hat{E}$ with the $z$-th component of a spin $\hat{\vec{S}} = (\hat{S}^x,\hat{S}^y,\hat{S}^z)$ of length $|l| = \sqrt{l(l+1)}$,
\begin{equation}\label{eq:SpinTruncation}
    \hat{E} \longmapsto \hat{S}^z = \sum_{i=-l}^{l} i \lvert i \rangle \langle i \rvert.
\end{equation}
This opens two different paths to implement the truncation for $\hat{U}$. 
The first employs the spin lowering $\hat{S}^{-} = \hat{S}^{\textrm{x}} - i\hat{S}^{y}$ and raising $\hat{S}^{+} = \hat{S}^x + i\hat{S}^y$ operators, such that $\hat{U} \mapsto \hat{S}^{-}/|l|$. 
The second prescribes
\begin{gather}\label{eq:S-operator}
\hat{U} \longmapsto
\begin{bmatrix}
  0 &  \dots & \dots & 0 \\
  1 &  \dots & \dots & 0 \\
  0 &  \ddots & \vdots & 0 \\
  0 &  \dots & 1 & 0\\
\end{bmatrix}.
\end{gather}

For $|l| \rightarrow \infty$, both mappings for $\hat{U}$ ensure $\hat{U}^{\dagger}\hat{U} = 1$ and the correct commutation relations between $\hat{E}$ and $\hat{U}$. The errors introduced by finite $|l|$ have been studied in Refs.~\cite{verstraete2008matrix,paper1}, where it is proven that are negligible in most scenarios \footnote{We remark that mappings for the operator $\hat{U}$ that preserve unitarity are possible, as in Refs.~\cite{PhysRevLett.111.115303,Kuno_2015,PhysRevD.95.094507,PhysRevA.94.063641}}. The specific choice of the truncation scheme depends on the quantum hardware that is employed. As an example, for quantum systems that allow for the implementation of interacting spin chains with large spins, such as $l = 1$ \cite{senko2015realization}, the definition based on the spin lowering $\hat{S}^{-}$ and raising $\hat{S}^+$ is ideal. 
For qubit-based quantum hardware, it is convenient to employ Eq.~\eqref{eq:S-operator} for the representation of $\hat{U}$, as done in the following.\\
\\For a given $l$, each gauge degree of freedom is then described by the $2l+1$ states $\ket{e},~e = -l$, $-l+1, \dots, ~0, \dots, ~l-1, ~l$. We map this vector space onto $2l+1$ qubits using 
\begin{align} \label{eq:encodedstates}
  \ket{-l + j} &= \ket{ \overbrace{0 \hdots 0}^{j} 1 \overbrace{0 \hdots 0}^{2l-j}},
\end{align}
where $0 \leq j \leq 2l$. With this encoding, the gauge field operators might be replaced by the simple forms
\begin{subequations}\label{eq:encodedoperators}
  \begin{align}
    \hat{E}&\mapsto \hat{S}^z = \frac{1}{2} \sum_{i = 1}^{2l} \prod_{j=1}^{i} \hat{\sigma}_{j}^{z}, \\
    \hat{U}&\mapsto \sum_{i = 1}^{2l} \hat{\sigma}_{i}^{-} \hat{\sigma}_{i+1}^{+}, \label{eq:encodedoperatorsU}
  \end{align}
\end{subequations}
where $\hat{\sigma}_i^{\pm} = \frac{1}{2}(\hat{\sigma}_i^x \pm i\hat{\sigma}_i^{y})$, and $\hat{\sigma}_i^{\textrm{x}}$, $\hat{\sigma}_i^{y}$, $\hat{\sigma}_i^{z}$ are the Pauli operators associated with the $i^{\textrm{th}}$ qubit. 
From these mapping we directly recover the relations $\hat{E}\ket{e} = e\ket{e}$ and $\hat{U}\ket{e} = (1 - \delta_{e, -l})\ket{e-1}$ for all $-l \leq e \leq l$. As an example, for $l=1$ the states in the gauge field basis become
\begin{subequations} \label{eq:truncatedgaugestates}
\begin{align}
	\ket{1} = &~\ket{001}, \\
	\ket{0} = &~\ket{010}, \\
	\ket{-1} = &~\ket{100}.
\end{align}
\end{subequations}
\\For our protocol, the required resources scale linearly in terms of both the parameter $l$ and the number of gauge and matter fields.
The encoding presented in Eq.~\eqref{eq:encodedstates} requires $2l+1$ qubits for storing $2l+1$ states. In principle, the same information can be stored in $\log (2l+1)$ qubits \cite{lewis2019qubit, Shaw_2020}. 
The reasons for which we choose the qubit encoding in Eq.~\eqref{eq:encodedstates} are the following.
First, as demonstrated in Ref.~\cite{paper1}, small values of $l$ allow for a good description of the untruncated model. Second, the terms assembling the operator $\hat{U}$ in Eq.~\eqref{eq:encodedoperatorsU} are easily implementable on different hardware platforms and are even native in trapped ion systems in the form of MS-gates. Finally, the states in Eq.~\eqref{eq:encodedstates} form a subspace of fixed magnetization, which is decoherence-free under the action of correlated noise (e.g. a globally fluctuating magnetic field) and allows for detection of single qubit bit-flip errors. Importantly, the latter is separately true for each individual gauge field and for the fermionic state. Since all the utilized quantum states lay in the single excitation subspace, a measurement resulting in states outside of this subspace can only be due to errors occurring during the quantum evolution, as long as the applied gates conserve the excitations. Hence, erroneous outcomes can be discriminated from faithful ones.\\
\\In the following, we will use the term ``physical states" to refer to the computationally relevant qubit states. 
For the gauge fields, this means that the qubit states lie in the computational space spanned by the states in Eqs.~\eqref{eq:truncatedgaugestates}. 
For the matter fields whose computational states are obtained via the Jordan-Wigner transformation, we operate in the zero-charge subsector (see Sec.~\ref{subsec:encodedHamiltonian_LatticeQEDIn2+1Dimensions}), which translates into matter states of zero magnetization. As a final remark, we highlight that our encoding for the electric field operators $\hat{E}$ and $\hat{U}$ applies equally well to the rotator $\hat{R}$ and plaquette $\hat{P}$ operator used for the plaquette with PBC (see Sec.~\ref{subsec:encodedHamiltonian_effectiveHamiltonianForPeriodicBoundaryConditions}).
\subsection{Open boundary conditions: dynamically generated magnetic fields}
\label{subsec:trappedIonResults_openBoundaryConditions}
\begin{figure}
	\centering
	\includegraphics[width=\columnwidth]{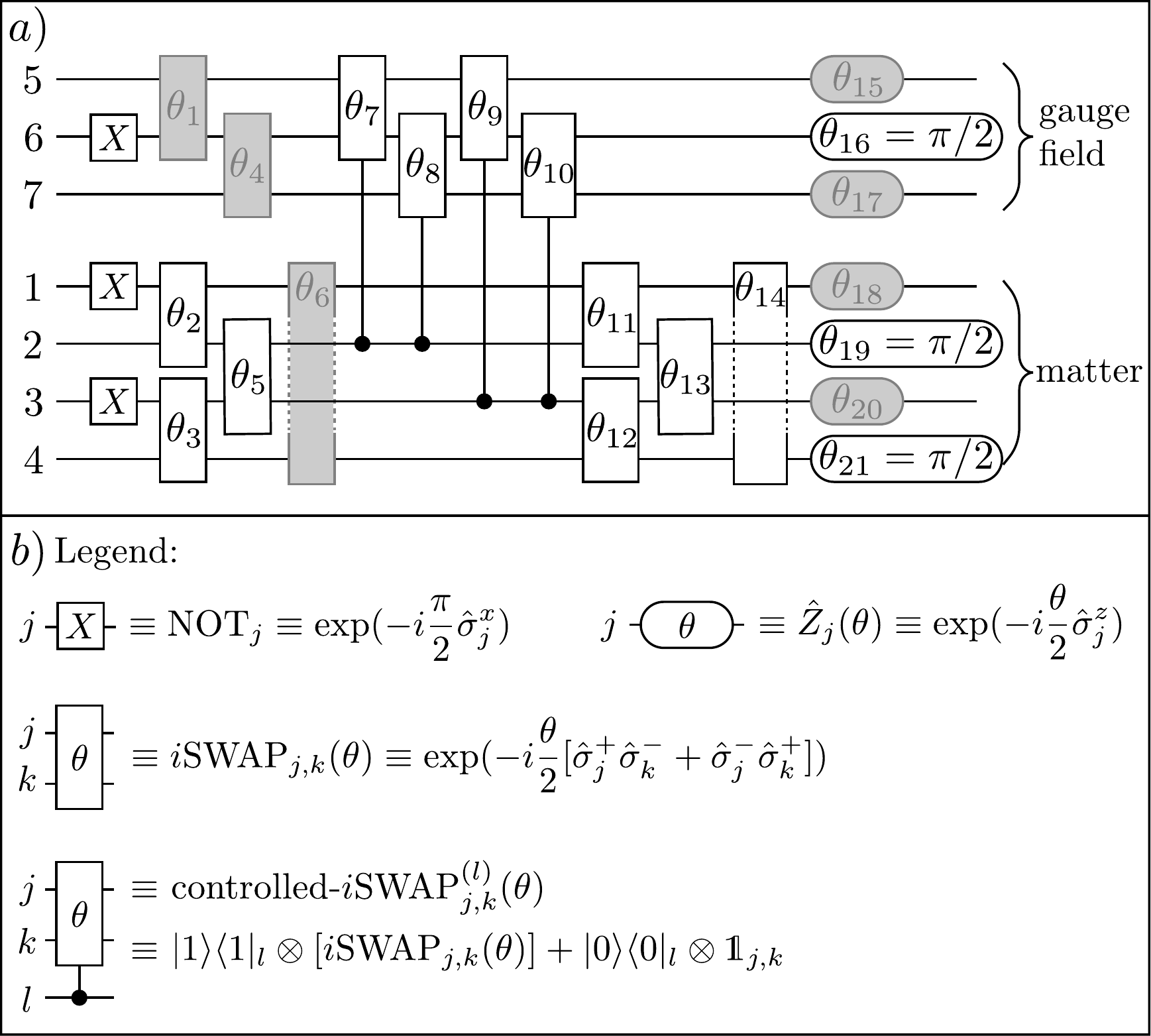}
	\caption{VQE circuit for preparing the ground state of a plaquette with open boundary conditions. \textbf{(a)} Quantum circuit with variational parameters $\theta_l$ shown for each gate. By identifying symmetries and redundancies via classical simulation, gates shaded in grey can be eliminated, and we set $\theta_{16} = \theta_{19} = \theta_{21} = \pi/2$, leaving eleven variational parameters. \textbf{(b)} Definitions of the gates used.}
	\label{fig:openCircuit}
\end{figure}
%
%
\noindent In the case of a plaquette with OBC, the simulation involves four qubits for the matter fields and $2l+1$ qubits for the gauge field. Here, we consider $l=1$ and thus the system consists of seven qubits. According to Fig.~\ref{fig:openCircuit}(c), we number the matter qubits as $1$, $2$, $3$, and $4$, and the gauge qubits $5$, $6$, and $7$. Plugging the encoding presented in Sec.~\ref{subsec:variationalQuantumSimulation_qubitEncoding} into the Hamiltonian in Eq.~\eqref{eq:singlePlaquetteHamiltonian}, we get
\begin{subequations} \label{eq:singlePlaquetteHamiltonianEncoded}
  \begin{align}
    \hat{H}_{\textrm{E}} = &~\frac{g^2}{4}\Big\lbrace \hat{\sigma}_{5}^z\left[ \hat{\sigma}_{1}^z-\hat{\sigma}_{3}^z + \hat{\sigma}_{6}^z(\hat{\sigma}_{1}^z - \hat{\sigma}_{3}^z - 2) - 1 \right] \nonumber \\
    & + \hat{\sigma}_{2}^z \left[ \hat{\sigma}_{1}^z + 2 \hat{\sigma}_{5}^z \left( \hat{\sigma}_{6}^z + 1 \right) - 1 \right] + 4\hat{\sigma}_{6}^z \Big\rbrace,  \\
    \hat{H}_{\textrm{B}} = &-\frac{1}{2g^2}\left[ \hat{\sigma}_{6}^+ \left( \hat{\sigma}_{5}^- + \hat{\sigma}_{7}^- \right) + \hat{\sigma}_{6}^- \left( \hat{\sigma}_{5}^+ + \hat{\sigma}_{7}^+ \right) \right], \\
    \hat{H}_{\textrm{m}} = &~\frac{m}{2} \Big(\hat{\sigma}_1^z - \hat{\sigma}_2^z + \hat{\sigma}_3^z - \hat{\sigma}_4^z\Big), \\
    \hat{H}_{\textrm{kin}} = &~-i\Omega \Big[\hat{\sigma}_{1}^+ \hat{\sigma}_{2}^- + \hat{\sigma}_{1}^+ \hat{\sigma}_{4}^-  + \hat{\sigma}_{4}^-\hat{\sigma}_{3}^+   \nonumber\\
    & - \hat{\sigma}_{2}^- \left( \hat{\sigma}_{5}^+ \hat{\sigma}_{6}^- + \hat{\sigma}_{6}^+ \hat{\sigma}_{7}^- \right) \hat{\sigma}_{3}^+ \Big] + \textrm{H.c.}. \label{eq:singlePlaquetteHamiltonianHkinEncoded}
  \end{align}
\end{subequations}
%
The matter qubit states are also given in Fig.~\ref{fig:openBoundaryConventions}(c). 
Recall that we chose the encoding such that physical states have total magnetization $\expval{\hat{S}_{\textrm{tot}}^{z}} = 1$. \\
\\The VQE quantum circuit shown in Fig.~\ref{fig:openCircuit}(a) preserves not only the total magnetization of the system, but also the magnetization of each of the gauge and matter subsystems. Hence, as mentioned in Sec.~\ref{subsec:variationalQuantumSimulation_qubitEncoding}, our magnetization-preserving quantum circuit used in combination with physical input states confines the VQE to the space of physical states.\\
\\The VQE circuit in its unreduced form [i.e., including the grey-shaded gates in Fig.~\ref{fig:openCircuit}(a)] is motivated by the form of the Hamiltonian in Eqs.~\eqref{eq:singlePlaquetteHamiltonian}. 
All qubits are initialized in the input state $\ket{0}$, and NOT gates prepare the bare vacuum $\ket{vvvv}\ket{0}$ [see Fig.~\ref{fig:openBoundaryConventions}(c)] as the initial state for the VQE. 
For the gauge field subsystem, the application of the parameterized $i$SWAP gates allows for accessing all three gauge field states $\ket{1}$, $\ket{0}$, and $\ket{-1}$, ensuring that the free ground state for $\Omega=0$ could be produced. 
Similarly, the parameterized $i$SWAP gates on the qubits $1$ to $4$ are used to allow for all physical basis states within the matter subsystem, and resemble the hopping terms of the kinetic Hamiltonian in Eq.~\eqref{eq:singlePlaquetteHamiltonianHkinEncoded}. These gates correspond to particle-antiparticle pair creation/annihilation in the model, and as a consequence, all matter basis states in the zero-charge subsector are made available by this part of the circuit. 
The kinetic Hamiltonian is likewise responsible for the entanglement between the subsystems, as the pair creation/annihilation processes are combined with a correction of the gauge field. 
The layer of parameterized controlled-$i$SWAP gates hence takes the role of the annihilation operator $\hat{U}$ and entangles the matter and gauge subsystems.
In the effective Hamiltonian of Eqs.~\eqref{eq:singlePlaquetteHamiltonian}, the gauge degree of freedom lies on the $(2,3)$ link and is directly coupled to matter sites $2$ and $3$. Accordingly, the circuit couples the gauge field with only these two fermions, which act as controls in the layer of controlled-$i$SWAP gates. 
Finally, parameterized $i$SWAP gates are applied on the matter qubits again to adjust the state after entangling the two subsystems, and a layer of single-qubit $z$-rotations is utilized to correct for relative phases. Other single-qubit operations are avoided as they are generally not magnetization preserving. We highlight that this circuit (as well as the ones in Fig.~\ref{fig:periodicCircuit}), being Hamiltonian inspired, from one side ensures the capability of exploring the subsector of physical states in the Hilbert space. From the other, it avoids redundant gates and limits the circuit depth, ensuring that barren plateaus \cite{mcclean2018barren,Uvarov_2020} in the energy landscape are prevented \cite{cerezo2021cost} when increasing the truncation $l$ and/or the number of plaquettes.\\
\begin{figure}
	\includegraphics[width=\columnwidth]{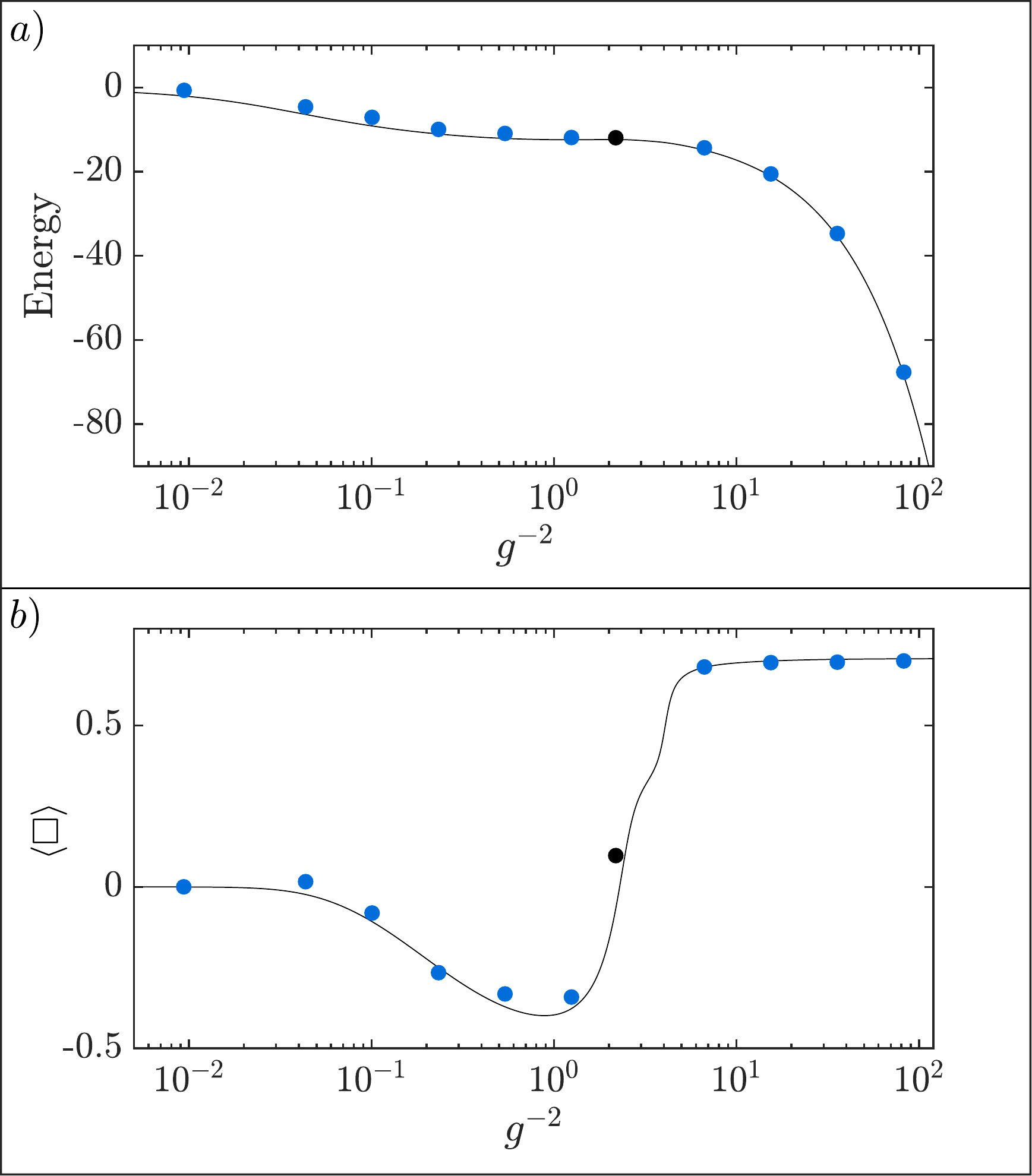}
	\caption{Classical simulation of the proposed experiment for observing the dynamical generation of magnetic fields where $\Omega = 5$ and $m = 0.1$ (see Sec.~\ref{subsec:quantumSimulation_openBoundaryConditions}) using the circuits given in Fig.~\ref{fig:openCircuit}. The blue and black dots represent data points obtained by variational minimization with a total finite measurement budget of $10^7$ measurements for the entire plot. Half of this budget is spent for the black dot alone. The black solid lines are determined via exact diagonalization of the Hamiltonians in Eqs.~\eqref{eq:singlePlaquetteHamiltonianEncoded}. \textbf{(a)} Energy of the variational ground state. \textbf{(b)} Plaquette expectation value $\expval{\Box}$ as a function of $g^{-2}$. All dots are calculated using the exact state corresponding to the optimal variational parameters found by the VQE.}
	\label{fig:openResults}
\end{figure}
\\The quantum circuit described above involves a total of $21$ variational parameters. While not strictly necessary, it is beneficial for currently available quantum hardware to reduce the number of variational parameters. This can be done by identifying the intrinsic symmetries of the ground state. By classically simulating the circuit in Fig.~\ref{fig:openCircuit}, we find that the solution space is still accessible by fixing $\theta_{16} = \theta_{19} = \theta_{21} = \pi/2$, and removing the gates shaded in grey in Fig.~\ref{fig:openCircuit}(a), leaving a total of eleven parameters. Furthermore, $\theta_{11}$, $\theta_{12}$ and $\theta_{13}$ can be set to zero outside the transition region $2 \lesssim g^{-2} \lesssim 5$, within which an almost vanishing energy gap between the ground and first excited state requires more precision for correctly estimating the ground state.\\
\\Since the circuit design is based on the structure of the Hamiltonian, the same design principles can be applied to larger scale systems. When adding more plaquettes, additional parametric $i$SWAP gates are used to populate all basis states within the matter and gauge subsystems. Then, the gauge degrees of freedom are coupled to their respective neighbouring matter sites using additional controlled-$i$SWAP gates [in correspondence with Eq.~\eqref{eq:OpenBoundarySpinHamiltoniankin}]. Finally, $z$-rotations adjust the relative phases of all qubits. When increasing the truncation $l$, additional $i$SWAP gates are inserted for populating the newly introduced gauge field states, and controlled-$i$SWAP gates are added for entangling them with the respective matter sites. In both cases -- adding more plaquettes and increasing the truncation cut-off -- a linear increase in the number of qubits and $i$SWAP gates and a quadratic increase in the number of controlled-$i$SWAP operations is expected.\\
\\A classical simulation of the proposed experiment, including statistical noise on the cost function $\mathcal{C}(\boldsymbol{\theta})$ (representative of the probabilistic nature of quantum state measurements -- see App.~\ref{app:optimalPartitioning}), is shown in Fig.~\ref{fig:openResults}. The data points (blue and black dots) correspond to the lowest energies found by the VQE. The energies and the plaquette expectation values are calculated using the exact state corresponding to the optimal variational parameters found for each value of $g^{-2}$. We verified that the VQE resorts to statistical errors affecting $\mathcal{C}(\boldsymbol{\theta})$ that are always lower than the ground and first excited states' energy gap. The black solid lines are obtained via exact diagonalization of the Hamiltonian in Eqs.~\eqref{eq:singlePlaquetteHamiltonianEncoded}. Using the measurement procedure described in App.~\ref{app:optimalPartitioning} and taking statistical error into account, the entire plot corresponds to approximately $10^7$ measurements to be performed on the quantum device. Half of this budget is used for the point at $g^{-2} \simeq 2.18$, where the energy difference between the ground and first excited states is much smaller if compared to other values of $g^{-2}$ (see above).\\
\\The ground state energy found by the VQE approximates well the energy of the exact ground state, as shown in Fig.~\ref{fig:openResults}(a). The plaquette expectation value is sensitive to small changes in the variational state, which leads to relatively large deviations with respect to the results obtained via exact diagonalization in Fig.~\ref{fig:openResults}(b), even for states whose energy is very close to the exact energy. Yet, the variational optimization is able to accurately resolve transitions in the order parameter. The fidelity of the variational ground state with respect to the exact ground state is particularly high in the extremal regions, exceeding $98\%$. All points achieve a fidelity greater than $90\%$.\\
\subsection{Periodic boundary conditions: running coupling}
\label{subsec:trappedIonResults_periodicBoundaryConditions}
\begin{figure}
	\centering
	\includegraphics[width=\columnwidth]{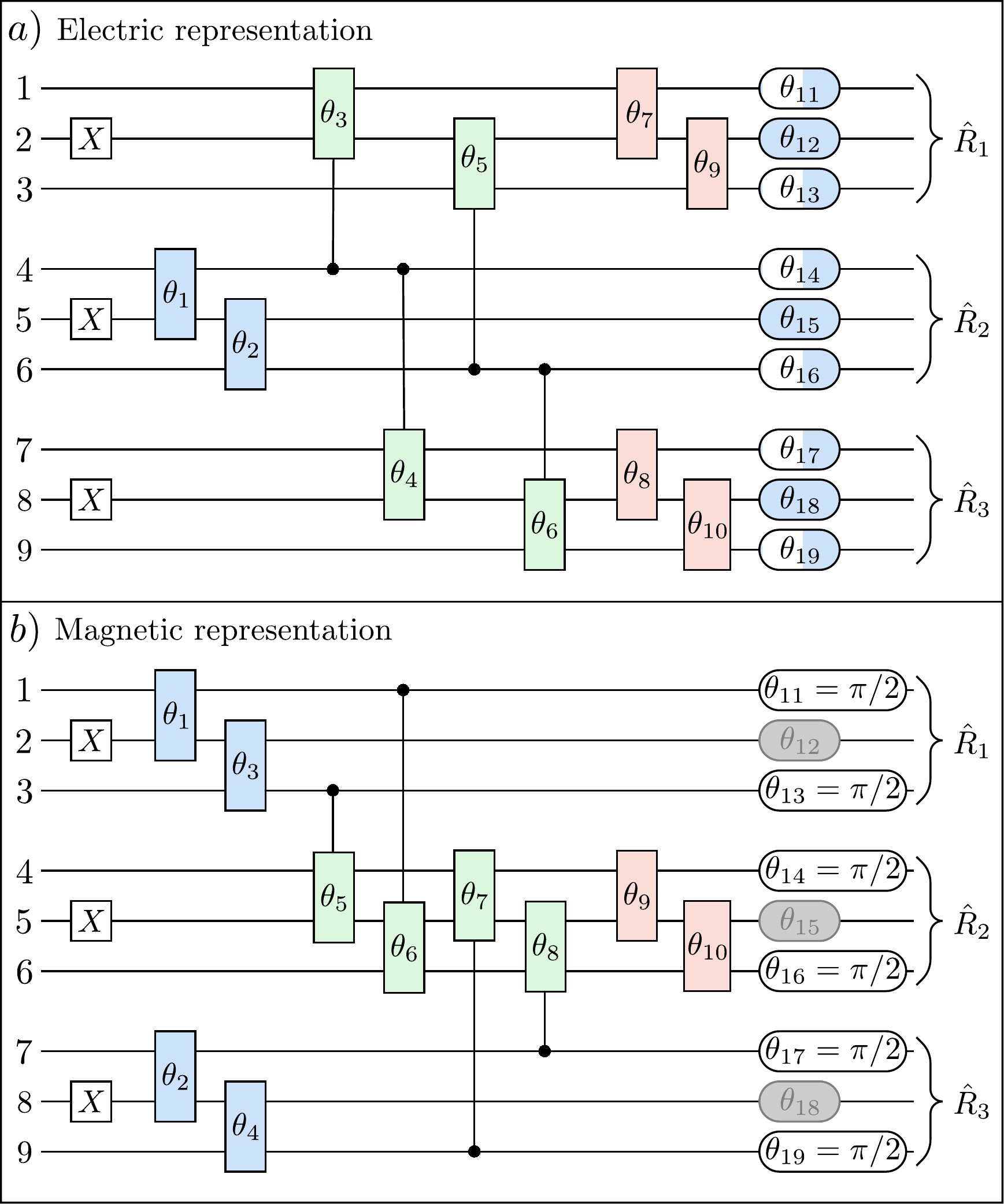}
	\caption{VQE circuits for preparing the ground state of a plaquette with periodic boundary conditions. Gate definitions are given in Fig.~\ref{fig:openCircuit}. Gates of the same color in each circuit share a variational parameter after eliminating redundant parameters. The half-shaded gates include an offset of $+\pi/2$ added to the shared parameter, while the grey-shaded gates can be eliminated entirely. \textbf{(a)} Circuit for the electric representation of the Hamiltonian, and \textbf{(b)} for the magnetic representation.}
	\label{fig:periodicCircuit}
\end{figure}
\begin{figure}
	\includegraphics[width=\columnwidth]{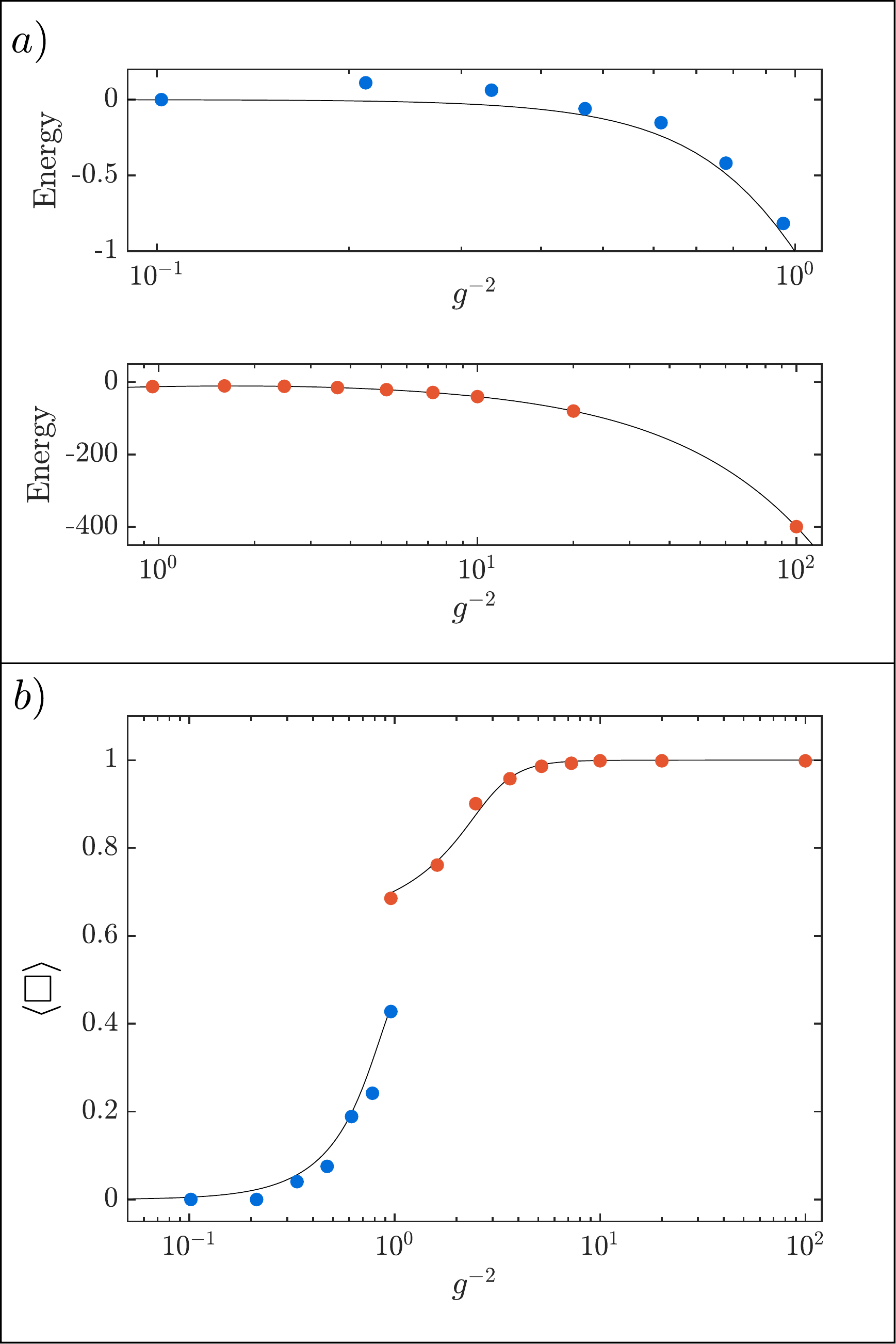}
	\caption{Classical simulation of the proposed experiment for observing the running of the coupling (see Sec.~\ref{subsec:quantumSimulation_periodicBoundaryConditions}) using the circuits given in Fig.~\ref{fig:periodicCircuit}. The red and blue data points are obtained from variational minimization with a total finite measurement budget of $6 \times 10^5$ measurements for the entire plot. The electric (magnetic) representation is shown in blue (red), and the black solid lines are determined via exact diagonalization of the Hamiltonians in Eqs.~\eqref{eq:periodicHamiltonianElectric} and \eqref{eq:periodicHamiltonianMagnetic}. \textbf{(a)} Energy of the variational ground state using the electric representation in the region $g^{-2} < 1$ and using the magnetic representation in the region $g^{-2} > 1$. \textbf{(b)} Plaquette expectation value $\expval{\Box}$ as a function of $g^{-2}$. All dots are calculated using the exact state corresponding to the optimal variational parameters found by the VQE.}
	\label{fig:periodicResults}
\end{figure}

\noindent In this section, we provide a VQE protocol for simulating the running coupling in LGTs. As explained in Sec.~\ref{subsec:quantumSimulation_periodicBoundaryConditions}, the running coupling is a genuine 2D effect that can be studied experimentally in a proof-of-concept demonstration by first preparing the ground state of a plaquette with PBC and subsequently measuring the expectation value $\langle \Box \rangle$.\\
\\Differently from the plaquette with OPC, the electric and magnetic representations of the Hamiltonian [see Eqs.~\eqref{eq:periodicHamiltonianElectricEncoded} and \eqref{eq:periodicHamiltonianMagneticEncoded}] are used for different regions of the bare coupling $g^{-2}$. This is done to obtain better convergence to the untruncated result, as the two representations are well suited for the strong and weak coupling regimes, respectively (see Sec.~\ref{subsec:encodedHamiltonian_effectiveHamiltonianForPeriodicBoundaryConditions} and Ref.~\cite{paper1}). Noting that the definitions in Sec.~\ref{subsec:variationalQuantumSimulation_qubitEncoding} presented for the electric gauge field and their lowering operators are trivially extended to rotators and plaquette operators, we encode the plaquette with PBC into nine qubits. Rotator $1$ is represented by qubits $1$ through $3$, rotator $2$ by qubits $4$ through $6$, and rotator $3$ by qubits $7$ through $9$ (see Fig.~\ref{fig:periodicCircuit}). Thus, the Hamiltonian becomes
\begin{subequations} \label{eq:periodicHamiltonianElectricEncoded}
  \begin{align}
    \hat{H}_{\textrm{E}}^{(e)} = & -\frac{g^2}{2} \Big\lbrace \hat{\sigma}_4^z\left( \hat{\sigma}_5^z + 1 \right)\left( \hat{\sigma}_1^z + \hat{\sigma}_7^z + \hat{\sigma}_7^z \hat{\sigma}_8^z \right) \nonumber\\& + \hat{\sigma}_2^z \left[ \hat{\sigma}_1^z \hat{\sigma}_4^z \left( \hat{\sigma}_5^z + 1 \right) -2 \right] -2 \left( \hat{\sigma}_5^z + \hat{\sigma}_8^z + 3 \right)  \Big\rbrace, \\
     \hat{H}_{\textrm{B}}^{(e)} = &-\frac{1}{2g^2}\Big[ \hat{\sigma}_1^- \hat{\sigma}_2^+ +\hat{\sigma}_2^- \hat{\sigma}_3^+ + \hat{\sigma}_4^- \hat{\sigma}_5^+ + \hat{\sigma}_5^- \hat{\sigma}_6^+  \nonumber\\& + \hat{\sigma}_7^- \hat{\sigma}_8^+ + \hat{\sigma}_8^- \hat{\sigma}_9^+ + \left( \hat{\sigma}_1^- \hat{\sigma}_2^+ + \hat{\sigma}_2^- \hat{\sigma}_3^+ \right)   \nonumber\\& \times \left( \hat{\sigma}_4^- \hat{\sigma}_5^+ + \hat{\sigma}_5^- \hat{\sigma}_6^+ \right) \left( \hat{\sigma}_7^- \hat{\sigma}_8^+ + \hat{\sigma}_8^- \hat{\sigma}_9^+ \right)  \Big] + \textrm{H.c.}
  \end{align}
\end{subequations}
in case $g^{-2} \lesssim 1$, and
\begin{subequations} \label{eq:periodicHamiltonianMagneticEncoded}
  \begin{align}
    \hat{H}_{\textrm{E}}^{(b)} = &~g^2 \sum_{\nu = 1}^{2L} \Bigg\{f_{\nu}^c \sum_{i=1}^{3} \left( \hat{\sigma}_{3i-2}^- \hat{\sigma}_{3i-1}^+ + \hat{\sigma}_{3i-1}^- \hat{\sigma}_{3i}^+ \right)^{\nu} \nonumber\\
		& + \frac{f_{\nu}^{s}}{2}\left[\left( \hat{\sigma}_{4}^- \hat{\sigma}_{5}^+ + \hat{\sigma}_{5}^- \hat{\sigma}_{6}^+ \right)^{\nu} - \left( \hat{\sigma}_{4}^+ \hat{\sigma}_{5}^- + \hat{\sigma}_{5}^+ \hat{\sigma}_{6}^- \right)^{\nu}\right] \nonumber\\
		& \times \sum_{\mu = 1}^{2L} f_{\mu}^{s} \Big[\left(\hat{\sigma}_{1}^- \hat{\sigma}_{2}^+ + \hat{\sigma}_{2}^- \hat{\sigma}_{3}^+ \right)^{\mu} \nonumber\\
		& + \left(\hat{\sigma}_{7}^- \hat{\sigma}_{8}^+ + \hat{\sigma}_{8}^- \hat{\sigma}_{9}^+ \right)^{\mu}\Big]\Bigg\} + \textrm{H.c.}, \\
    \hat{H}_{\textrm{B}}^{(b)} = &-\frac{1}{g^2} \Bigg[ \sum_{i = 1}^{3} \cos \left(\frac{\pi \left( \hat{\sigma}_{3i-2}^z + \hat{\sigma}_{3i-2}^z \hat{\sigma}_{3i-1}^z \right)}{2L+1}\right)  \nonumber \\
    & + \cos \left(\frac{\pi \sum_{i=1}^3 \left(\hat{\sigma}_{3i-2}^z + \hat{\sigma}_{3i-2}^z \hat{\sigma}_{3i-1}^z \right)}{2L+1}\right)\Bigg]
  \end{align}
\end{subequations}
whenever $g^{-2} \gtrsim 1$. \\
\\As for the case of OBC, the circuit design for a plaquette with PBC is motivated by the structure of the Hamiltonian and employs the same gate set. Due to the differences between the electric and magnetic representations, we use two different VQE circuits which are shown in Fig.~\ref{fig:periodicCircuit}. 
Contrary to the plaquette with OBC, the controlled $i$SWAP gate are not used to entangle the matter with the gauge subsystems. Instead, they are motivated by the coupling of the rotators and plaquette operators in the last terms of Eqs.~\eqref{eq:periodicHamiltonianElectricE} and \eqref{eq:periodicHamiltonianElectricB}, respectively, and the corresponding ones in the magnetic representation.\\
\\Both circuits have $19$ variational parameters and allow us to thoroughly explore the associated Hilbert spaces. However, since we are solely interested in the system ground state, it is convenient for NISQ technology to reduce the number of variational parameters by exploiting the symmetries between rotators $1$ and $3$ [which are apparent from the Hamiltonians in Eqs.~\eqref{eq:periodicHamiltonianElectric} and \eqref{eq:periodicHamiltonianMagnetic}] and by identifying additional redundant parameters through classical simulation of the VQE. As a result, in Fig.~\ref{fig:periodicCircuit}(a) we use a single parameter for each of the following sets: $\{\theta_{1},\theta_{2}, \theta_{11}, \dots, \theta_{19}\}$, $\{\theta_{3} ,\dots \theta_{6}\}$, and $\{\theta_{7},\dots, \theta_{10}\}$, as indicated by the color coding. Parameters $\theta_{11}$, $\theta_{13}$, $\theta_{14}$, $\theta_{16}$, $\theta_{17}$, and $\theta_{19}$ include an offset of $+\pi/2$ added to the shared variational parameter, which is indicated by the half-shaded gates.
For the circuit in Fig.~\ref{fig:periodicCircuit}(b), we make the groupings $\{\theta_{1}, \theta_{2}, \theta_{3},\theta_{4}\}$, $\{\theta_{5} ,\dots, \theta_{8}\}$, and $\{\theta_{9}, \theta_{10}\}$, while $\theta_{12}$, $\theta_{15}$, and $\theta_{18}$ can be eliminated, and $\theta_{11}$, $\theta_{13}$, $\theta_{14}$, $\theta_{16}$, $\theta_{17}$, and $\theta_{19}$ are fixed at $\pi/2$. This leaves just three variational parameters for each circuit.\\
\\The circuit for the electric representation of Eqs.~\eqref{eq:periodicHamiltonianElectricEncoded} is shown in Fig.~\ref{fig:periodicCircuit}(a). All qubits are initialized in the input state $\ket{0}$, and NOT gates prepare the vacuum state $\ket{0} = \ket{010}$ for each of the three rotators as the intial state for the VQE. The layer of controlled-$i$SWAP gates reflects the coupling between the rotators in the electric Hamiltonian $\hat{H}_{\textrm{E}}^{(e)}$ [see Eq.~\eqref{eq:periodicHamiltonianElectricE}], which takes the form $-\hat{S}^z_2 (\hat{S}^z_1 + \hat{S}^z_3)$. This term results from the elimination of rotator $4$ as a redundant degree of freedom, and introduces an asymmetry between rotator $2$ and rotators $1$ and $3$. When increasing $g^{-2}$, the ground state spreads from $\ket{0} \ket{0} \ket{0}$ (in the rotator basis) to all other electric levels, and states in which all three rotators have the same sign ($\ket{1}\ket{1}\ket{1}$ and $\ket{-1} \ket{-1} \ket{-1}$) receive the strongest negative contribution. To encourage the VQE to prepare the correct superposition of states, the parameterized controlled-$i$SWAP gates are connected to control the spread of population within rotators $1$ and $3$ based on the population of rotator $2$.\\
\\The circuit for the magnetic representation of Eqs.~\eqref{eq:periodicHamiltonianMagneticEncoded} is shown in Fig.~\ref{fig:periodicCircuit}(b). Its construction is similar to the circuit for the electric representation described above, but rather encourages the flip-flop interactions between rotators described by $\hat{H}_{\textrm{E}}^{(b)}$ [see Eq.~\eqref{eq:periodicHamiltonianMagneticE}]. Both the electric and magnetic circuits maintain constant magnetization of each gauge field, which prevents access to unphysical states.\\
\\Designing the VQE circuit based on the form of the Hamiltonian allows for a scalable architecture. For systems with additional plaquettes, the coupling between rotators remains pairwise, which translates into the addition of controlled-$i$SWAP gates between all pairs of coupled gauge fields, as was described above for OBC. When considering larger truncations $l$, additional $i$SWAP gates are introduced to allow for all gauge field basis states. Additional controlled-$i$SWAP gates are then added to share entanglement in a similar fashion as for the case $l=1$ considered here. In both cases, the scaling is the same as for the OBC circuit, i.e., the number of qubits and the number of $i$SWAPs scale linearly, while the number of controlled-$i$SWAP gates scales quadratically in the worst-case scenario.\\
\\We simulate the proposed experiment classically, including statistical noise on the cost function $\mathcal{C}(\boldsymbol{\theta})$. Our results are shown in Fig.~\ref{fig:periodicResults}. Points obtained with the electric and the magnetic representation of the Hamiltonian are shown in blue and red, respectively, while the black solid lines come from exact diagonalization of the Hamiltonians. As for the OBC, the energy and the plaquette expectation value $\langle \Box \rangle$ are calculated using the exact state obtained with the optimal variational parameters found by the VQE. Using the measurement procedure described in App.~\ref{app:optimalPartitioning} and taking statistical errors into account, the entire plot corresponds to $6\times10^5$ measurements to be performed on the quantum device.\\
\\The VQE protocol reaches the correct ground state energy [see Fig.~\ref{fig:periodicResults}(a)], and the expectation value of the plaquette operator is accurate if compared to the exact truncated results [Fig.~\ref{fig:periodicResults}(b)]. The fidelity of the variational ground state with respect to the exact ground state exceeds $96\%$ for all points, and for the majority of points it exceeds $99\%$.\\
\subsection{Errors in ion platforms}
\label{subsec:trappedIonResults_errors}
\noindent In this section, we discuss the effects of experimental imperfections, as well as statistical noise, on our protocols. Since the results in Secs.~\ref{subsec:trappedIonResults_openBoundaryConditions} and \ref{subsec:trappedIonResults_periodicBoundaryConditions} are derived assuming an ion-based quantum computer \cite{zhang2017observation,shehab2019noise,RevModPhys.93.025001,kim2010quantum,rajabi2019dynamical,chertkov2021holographic}, we consider here the main sources of errors in such platforms. Our considerations are based on the experimental apparatus in Refs.~\cite{kokail2019self,marciniak2021optimal}, that has been previously used for LGT simulations in 1D. If the reader is interested in using a superconducting device, we refer to Refs.~\cite{arute2019quantum, corcoles2015demonstration,PhysRevLett.122.080504}.\\
\\In the following, we distinguish between extrinsic and intrinsic errors. Extrinsic errors are determined by the characteristics of the experimental apparatus and include imperfect gate operations, dephasing, and systematic errors such as offsets in the variational parameters $\boldsymbol{\theta}$. By contrast, intrinsic errors are inherent to the VQE protocol and as such unavoidable. These are statistical errors, and follow from the stochastic nature of quantum measurements (see App.~\ref{app:optimalPartitioning}).\\
\\As discussed in Refs.~\cite{kokail2019self,marciniak2021optimal}, for low circuit depths intrinsic errors are the dominant source of noise. Since the cost function $\mathcal{C}(\boldsymbol{\theta})$ is estimated from independent contributions (see App.~\ref{app:optimalPartitioning}), 
 we add up the uncertainties due to each term for separately. This substantially increases the uncertainty of $\mathcal{C}(\boldsymbol{\theta})$, even for eigenstates of the considered Hamiltonian. As an example, for each value of the cost function, obtaining a statistical error smaller than the energy gap between the ground and first excited state requires averaging several thousands of experimental shots. This is expected to be the main, and largest overhead from noise sources in our proposal.\\
\\
 Extrinsic errors have limited impact for the small system sizes considered here. Indeed, VQE schemes are robust against systematic over- or under-rotation in one- and two-qubits gates \cite{mcclean2016theory}. This follows from the fact that the classical optimizer determines the variational parameters $\boldsymbol{\theta}$ by minimizing the cost function. As such, since the shifts in $\boldsymbol{\theta}$ are slowly varying if compared to the time required for the cost function minimization, these errors are compensated by the classical subroutine. Furthermore, imperfect gates and finite coherence time limit the number of gates and the circuit depth that can be used, respectively. This results in a limitation in the number of plaquettes and truncation $l$ that can be studied. Such limitation can however be relaxed by technological improvements.\\
\section{Analytical interpretation of the results}
\label{sec:quantumSimulationOf2DEffects}
\noindent In this section, we use the effective Hamiltonian description provided in Sec.~\ref{sec:encodedHamiltonian} to study lattice QED with OBC (Sec.~\ref{subsec:quantumSimulation_openBoundaryConditions}) and PBC (Sec.~\ref{subsec:quantumSimulation_periodicBoundaryConditions}). While next generation quantum computers will widen the scope of our approach, we focus on phenomena that can be studied with small system sizes and for which quantum simulations can be carried out on current quantum hardware \cite{martinez2016real, nam2020ground, arute2019quantum, corcoles2015demonstration, bernien2017probing, labuhn2016tunable}. Here, we motivate the relevance of the results of the quantum simulations in Sec.~\ref{sec:main_result}.
\subsection{Dynamical matter and magnetic fields}
\label{subsec:quantumSimulation_openBoundaryConditions}
\begin{figure}
	\includegraphics[width=\columnwidth]{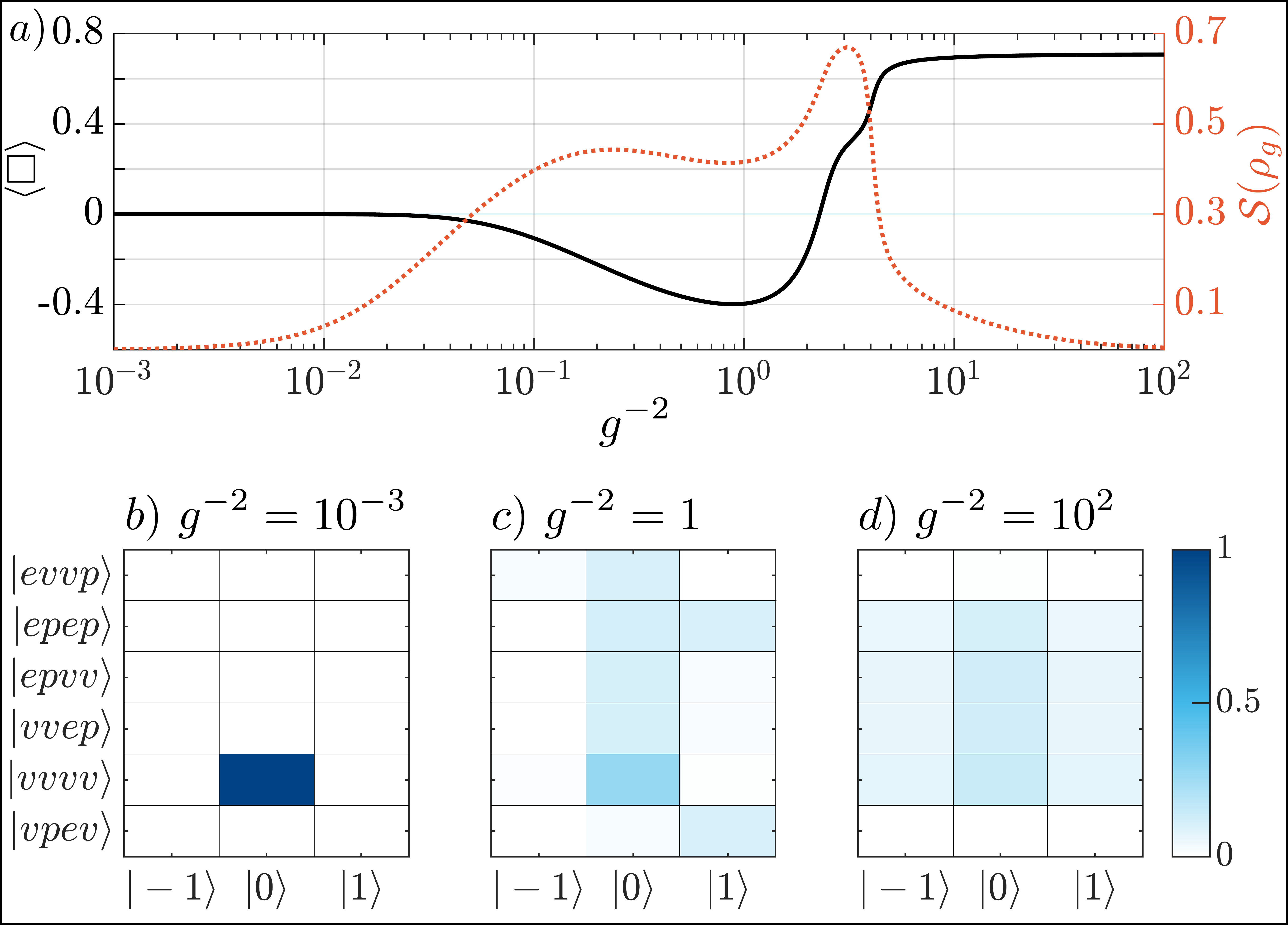}
	\caption{Ground state properties in the non-perturbative regime. \textbf{(a)} Plaquette operator expectation value $\expval{\Box}$ and entanglement entropy $\mathcal{S}(\rho_g)$ as a function of $g^{-2}$. \textbf{(b)-(d)} The probability by component of the ground state for the indicated $g^{-2}$. At $g^{-2} = 10^{-3}$, the lowest energy state is the vacuum $\ket{vvvv}\ket{0}$. For $g^{-2} = 1$, the ground state approximates that of the kinetic term. For large $g^{-2}$, the magnetic Hamiltonian in the presence of the kinetic Hamiltonian yields the tensor product between a superposition of matter states and the gauge ground state of the magnetic term. Through (a) - (d), we used $\Omega = 5$ and $m = 0.1$, and a gauge field truncation $l =1$.}
	\label{fig:plaquetteNonPerturbative}
\end{figure}

\noindent We describe a minimal example that allows for the study of 2D effects in lattice QED with OBC. More specifically, we examine the appearance of dynamically generated magnetic fields due to particle-antiparticle creation processes within a single plaquette. This effect manifests itself in an abrupt change of the ground state as a result of the competition between the kinetic and magnetic terms in the QED Hamiltonian of Eqs.~\eqref{eq:singlePlaquetteHamiltonian}. There are two parameter regimes in which this phenomenon occurs. In one, the parameter $\Omega$ [see Eq.~\eqref{eq:singlePlaquetteHamiltonianHkin}] is dominant if compared to the mass and the bare coupling $g$, allowing the kinetic and the magnetic terms to be the leading contributions to the energy. Alternatively, the ground state's shift appears when the mass is smaller than zero (or with positive mass and non-zero background field). This last scenario is hard to simulate using MCMC methods, in which the use of the inverse of the lattice Dirac operator in combination with a negative mass induces zero modes \cite{Gattringer:2010zz}, leading to unstable simulations. We note that in 1D QED, considering a negative fermion mass is equivalent to a theory with positive mass in the presence of a topological term $\theta = \pi$ (see, e.g., \cite{funcke2020topological}) \footnote{In principle, the fermion mass can be written with a phase factor $m e^{i\theta}$. In 1D QED, the anomaly equation (see the appendix of Ref.~\cite{funcke2020topological}) can be used to eliminate this phase by a rotation, and it appears as the topological term in the gauge field. This means that by working with a negative fermion mass, the model is equivalent to a theory with a $\theta$ term at $\theta = \pi$.}. It will be interesting to investigate the relation of the negative fermion mass to a theory in the presence of a topological term in $2$ dimensions.\\
 \\The proposed experiment consists of preparing the ground states of Eqs.~\eqref{eq:singlePlaquetteHamiltonian} for different values of the coupling $g$, and subsequently measuring the magnetic field energy, which is proportional to $\expval{\Box}$ [see Eq.~\eqref{eq:PlaquetteOperator}]. In the strong coupling regime $g^{-2} \ll 1$, the magnetic field energy vanishes $\expval{\Box} = 0$, since the ground state approaches the bare vacuum $\ket{vvvv}\ket{0}$. For weak couplings, $g^{-2} \gg 1$, the magnetic term dominates and the ground state is a superposition of all electric field basis elements. As such, $\expval{\Box}$ converges to 1 when $g^{-2} \rightarrow \infty$. However, the truncation described in Sec.~\ref{subsec:encodedHamiltonian_LatticeQEDIn2+1Dimensions} bounds $\expval{\Box}$ to a smaller value \cite{paper1}.\\
\begin{figure}[t]
  \includegraphics[width=\columnwidth]{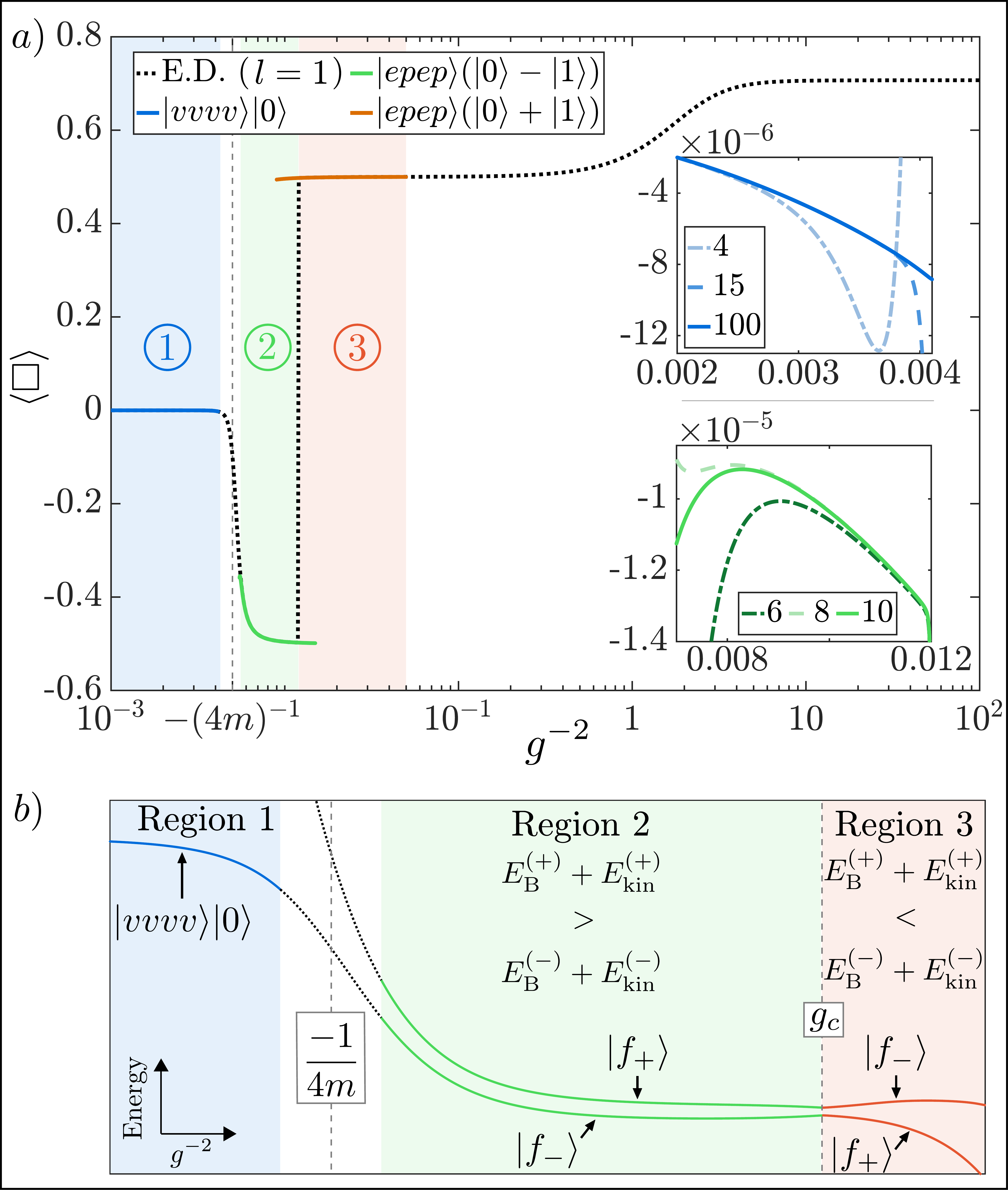}
  \caption{\textbf{(a)} Plaquette expectation value as a function of $g^{-2}$ for a single plaquette with mass $m = -50$ and kinetic strength $\Omega = 5$. Regions $1$, $2$, and $3$ correspond to the numbered regions in part (b) below. Exact diagonalization of the truncated model is represented by the black dashed line, and perturbation theory was used for the blue, green, and orange lines. The insets are magnifications of parts of regions 1 and 2 (blue and green, respectively), plotting the difference between exact diagonalization and perturbation theory of order reported in the legend to confirm convergence to the full U(1) theory. The difference is not completely vanishing due to the small contribution of the magnetic term which is ignored in perturbative calculations. \textbf{(b)} Schematic representation of the ground and first excited state's energy.}
  \label{fig:plaquettePerturbationPlot}
\end{figure}
\\To explain the underlying physics in the intermediate region between the strong and weak coupling regimes, we first examine the case in which the kinetic term of the Hamiltonian can be treated as a perturbation (Fig.~\ref{fig:plaquettePerturbationPlot}). Perturbative calculations are valid for $1 \ll \Omega \ll |4m|$, and we show results for $\Omega = 5$ and $m = -50$. We then proceed to the non-perturbative case (Fig.~\ref{fig:plaquetteNonPerturbative}) with parameters $\Omega = 5$ and $m = 0.1$. Our analytical results within the perturbative regime are determined using our iterative algorithm described in App.~\ref{app:PerturbationTheory}, and are unaffected by truncation effects in the parameter $l$. Therefore, they also serve as a test of validity for exact results.\\
\\For $\Omega = 5$ and $m = -50$, the kinetic term can be treated as a perturbation, and we take the sum of the electric and mass terms in Eqs.~\eqref{eq:singlePlaquetteHamiltonianHE} and \eqref{eq:singlePlaquetteHamiltonianHm} as the bare Hamiltonian. We can thus well describe the strong coupling regime $g^{-2} \ll 1$, in which a 2D effect occurs that is characterized by a jump of the expectation value of the plaquette operator. A plot of $\expval{\Box}$ as a function of $g^{-2}$ is shown in Fig.~\ref{fig:plaquettePerturbationPlot}(a), where the black dashed line is obtained using exact diagonalization of the Hamiltonian in Eqs.~\eqref{eq:singlePlaquetteHamiltonian} for a truncation $|l| = 1$, and the coloured lines correspond to perturbative calculations. We begin by analyzing the system for $g^{-2} \ll 1$ (marked as region $1$ in Fig.~\ref{fig:plaquettePerturbationPlot}), and examine increasing values of $g^{-2}$ as we move through regions $2$
and $3$.\\
\\In region $1$, the ground state of the Hamiltonian is essentially the bare vacuum $\ket{vvvv}\ket{0}$, which is characterized by null energy and $\expval{\Box} = 0$. The dominating electric term prevents the creation of electric field, despite the mass term incentivizing particle-antiparticle pair creation ($m <0$). The magnetic term is negligible in the strong coupling regime. With increasing $g^{-2}$, the cost of creating electric fields is reduced, until it becomes favourable to create particle-antiparticle pairs. In particular, for $g^{-2} = -(4m)^{-1}$, the energy relief $2m$ of creating a pair compensates the cost $g^2/2$ of creating an electric field on the link connecting the pair.\\
\\The kinetic term $\hat{H}_{\textrm{kin}}$ is responsible for the energy anti-crossing between regions $1$ and $2$ as it allows the creation of particle-antiparticle pairs and the corresponding electric fields. It couples the vacuum to the plaquette states which contains the maximum number of particle-antiparticle pairs (we refer to this as a fully-filled plaquette). However, Gauss' law allows two different gauge field configurations for the fully-filled plaquette, as shown in Fig.~\ref{fig:openBoundaryConventions}(e). These states are given by $\ket{f_{\pm}^{(0)}} = \frac{1}{\sqrt{2}}\ket{epep}(\ket{0} \pm \ket{1})$, and while they are degenerate with respect to the bare Hamiltonian $\hat{H}_{\textrm{E}} + \hat{H}_{\textrm{m}}$, $\expval{\hat{H}_{\textrm{kin}}}$ is minimized for $\ket{f_{-}}$ and $\expval{\hat{H}_{\textrm{B}}}$ for $\ket{f_{+}}$. The existence of these two configurations in the presence of the perturbation $\hat{H}_{\textrm{kin}}$ and the magnetic term $\hat{H}_{\textrm{B}}$ creates competition between two quasi-degenerate vacua in regions $2$ and $3$ of Fig.~\ref{fig:plaquettePerturbationPlot}(b).
The ground states in these regions are described by the corresponding corrected states in perturbation theory $\ket{f_{\pm}} = \sum_{n} \ket{f_{\pm}^{(n)}}$ (see App.~\ref{app:PerturbationTheory}), as shown in Fig.~\ref{fig:plaquettePerturbationPlot}(b). Here, we see that the kinetic term facilitates the creation of particle-antiparticle pairs and drives the ground state from the vacuum in region $1$ towards $\ket{f_{-}}$ in region $2$.\\
\\The remarkable 2D feature of the theory is the jump of $\expval{\Box}$ between regions $2$ and $3$, which is shown by the black dotted line in Fig.~\ref{fig:plaquettePerturbationPlot}(a). This jump corresponds to the sharp energy anti-crossing in Fig.~\ref{fig:plaquettePerturbationPlot}(b), and follows from the competition between the quasi-degenerate vacua in the presence of the kinetic and magnetic terms. As the relative weights of $\hat{H}_{\textrm{kin}}$ and $\hat{H}_{\textrm{B}}$ change with $g^{-2}$, an anti-crossing occurs and the ground state changes from $\ket{f_{-}}$ to $\ket{f_{+}}$. More specifically, there is a value $g_{\textrm{c}}$ such that, for any $g^{-2}  > g_{\textrm{c}}^{-2}$, $E_{\textrm{B}}^{(+)} + E_{\textrm{kin}}^{(+)} < E_{\textrm{B}}^{(-)} + E_{\textrm{kin}}^{(-)}$, where $E_{\alpha}^{(\pm)} = \bra{f_{\pm}} \hat{H}_{\alpha} \ket{f_{\pm}}$. Despite the fact that the magnetic Hamiltonian is not included in our perturbative analysis, we can analytically calculate this value $g_{\textrm{c}}$ (see App.~\ref{app:PerturbationTheory}).
By requiring the magnetic and kinetic contributions from the energy to be equal, we find $g_{\textrm{c}} = 0.012 + o(\Omega^8)$, which is in excellent agreement with the results obtained from exact diagonalization [see Fig.~\ref{fig:plaquettePerturbationPlot}(a)]. Following the jump, in region $3$, the ground state is thus $\ket{f_+}$, until we reach the weak coupling regime and the magnetic term becomes dominant compared to all other contributions.\\
\\The analysis above shows how pair creation processes can lead to dynamically generated magnetic fluxes that result in negative values of the magnetic field energy. For negative mass, the considered effect of competing vacua occurs over a wide range of parameters. The strength of the kinetic term $\Omega$ broadens the dip and shifts the position of the jump $g_{\textrm{c}}$.\\
\\In the following, we consider the regime in which the kinetic term cannot be treated as a perturbation. For $\Omega = 5$ and $m = 0.1$, exact diagonalization results are shown in Fig.~\ref{fig:plaquetteNonPerturbative}(a). In the intermediate region where $g \simeq 1$, the kinetic term has significant weight, which leads to entanglement between matter and gauge degrees of freedom in the ground state. In particular, in the limit $\Omega \gg |m|,g^{2},g^{-2}$ and for $l \rightarrow \infty$, it can be shown for $N$ plaquettes \cite{lieb1994flux} that the energy is minimized for a magnetic flux of $\pi$, which is generated by pair creation processes and corresponds to $\expval{\Box} = -1$ (which is not reached in Fig.~\ref{fig:plaquetteNonPerturbative}(a) due to the effects of truncation). Figures~\ref{fig:plaquetteNonPerturbative}(b)-(d) show the probabilities of the components of the ground state at different values of $g^{-2}$.
In the strong and weak coupling regimes, the ground states are the vacuum $\ket{vvvv}\ket{\textrm{GS}^{(e)}}$ and $\frac{1}{2}(\ket{vvvv}-\ket{epep} + i\ket{epvv} + i\ket{vvep}) \ket{\textrm{GS}^{(b)}}$, respectively, where $\ket{\textrm{GS}^{(e)}} = \ket{0}$ is the gauge component of the ground state of the electric term and $\ket{\textrm{GS}^{(b)}}$ is that of the magnetic term (see App.~\ref{app:newAppendix}). For $g^{-2}=1$, however, the ground state approximates the ground state of the truncated kinetic term for $l = 1$ with $92\%$ fidelity (which can be increased by incrementing $\Omega$). To quantify the ground state entanglement between matter and gauge degrees of freedom, we calculate the entanglement entropy $\mathcal{S}(\rho_g) = -\textrm{Tr}[\rho_g \log \rho_g]$, where $\rho_g$ is the density matrix of the gauge degrees of freedom that remain after eliminating those that are redundant. The entanglement entropy is plotted as the red dashed line in Fig.~\ref{fig:plaquetteNonPerturbative}(a).\\
\subsection{Running coupling}
\label{subsec:quantumSimulation_periodicBoundaryConditions}
\begin{figure}[t]
	\includegraphics[width=\columnwidth]{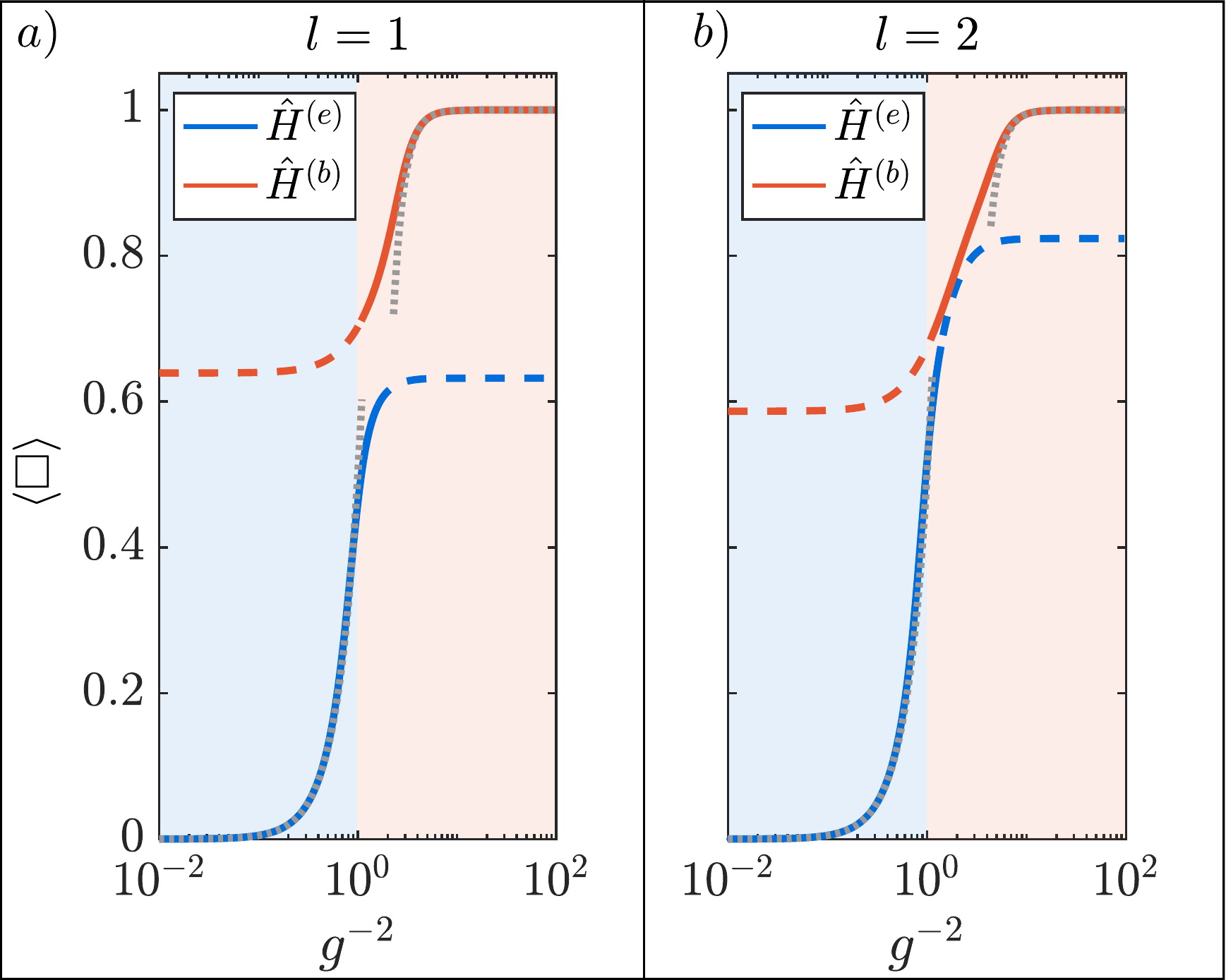}
	\caption{Ground state expectation value of the plaquette operator $\Box$ for a pure gauge periodic boundary plaquette (see Fig.~\ref{fig:periodicBoundaryConventions}). Results obtained using the electric and magnetic representation are shown as blue and red lines, respectively. Perturbative calculations are shown as grey dashed lines. The plots correspond to two different truncations $l=1$ and $l=2$, as indicated above. The blue (red) shaded region indicates the values of $g^{-2}$ for which the the electric (magnetic) representation achieves the best results.}
	\label{fig:plaquettePeriodicPlot}
\end{figure}

\noindent In this section, we consider 2D effects in QED on a lattice subject to PBC. The method described in Sec.~\ref{subsec:encodedHamiltonian_effectiveHamiltonianForPeriodicBoundaryConditions} and in Ref.~\cite{paper1} is based on the Hamiltonian formalism of LGTs, which allows for simulations that are unaffected by the problem of autocorrelations inherent to MCMC methods. Thus, we can compute physical observables at arbitrary values of the lattice spacing, ultimately allowing for reaching a well-controlled continuum limit. As such, future quantum hardware could be used to efficiently calculate, for instance, the bound state mass spectrum of the theory, properties of the inner structure of such bound states, or form factors which are important for experiments \cite{Gattringer:2010zz, aoki2020flag}.\\
\\We are therefore bound to consider local observables which only need a small number of lattice points.
We consider the plaquette operator $\Box$ as a simple example, as was done in the pioneering work by Creutz \cite{creutz1983monte} in the beginning of MCMC simulations
of LGTs. Despite its local nature, the operator $\Box$ can be related to a fundamental parameter of the theory, namely, the
renormalized coupling $g_{\rm ren}$ \cite{Booth:2001qp}. Importantly, the 1D Schwinger model is
super-renormalizable, meaning that the coupling does not get renormalized, while, as we will show, renormalization is necessary for our 2D model.\\
\\Renormalization appears in quantum field theories through quantum fluctuations, i.e., the spontaneous generation of particle-antiparticle pairs from the vacuum. This phenomenon leads to a charge shielding (or anti-shielding for non-Abelian gauge theories) \cite{Peskin:1995ev}, which in turn changes the strength of the charge depending on the distance (or the energy scale) at which it is probed. As such, the charge becomes {\em scale dependent} and in this work we choose the inverse lattice spacing $1/a$ as the scale. The scale dependence, which is a renormalization effect, is referred to as the {\em running} of the coupling and its knowledge is fundamental in understanding the interactions between elementary particles.
In particular, the running coupling serves as an input to interpret results from
collider experiments, such as the Large Hadron Collider. Hence, the ability to compute the running
coupling from the plaquette operator can have
a direct impact on such experiments and our knowledge of elementary particle interactions.\\
\\As explained in Sec.~\ref{subsec:encodedHamiltonian_effectiveHamiltonianForPeriodicBoundaryConditions}, we study the case of two-dimensional QED without matter. Despite the absence of matter, and in contrast to the previously studied 1D Schwinger model, renormalization is needed, and hence the calculation of the running coupling. A definition of the renormalized coupling $g_{\textrm{ren}}$ can be given through the ground state expectation value of the plaquette $\expval{\Box}$ \cite{Booth:2001qp}
\begin{equation}
g_{\rm ren}^2 = \frac{g^2}{\langle \Box \rangle^{1/4}},
\label{eq:plaquettecoupling}
\end{equation}
where $g$ is the bare coupling from the Hamiltonian [see Eqs.~\eqref{eq:periodicHamiltonianElectric} and \eqref{eq:periodicHamiltonianMagnetic}].
By looking at Eq.~\eqref{eq:plaquettecoupling}, it follows that to cover the scale dependence of the coupling, we need to evaluate $\expval{\Box}$ over a broad range of the lattice spacing $a$. Equivalently (we set $a=1$ above), this means that we need to perform simulations at many values of $g^{-2}$ , covering the whole spectrum between the extremal regions of strong and weak couplings, $g^{-2} \ll 1$ and $g^{-2} \gg 1$, respectively. In particular, the interesting region where perturbation theory is no longer applicable and bound states can be computed on not too large lattices is where $g^{-2} \simeq 1$. Covering such a broad range of scales is the major problem in standard lattice gauge simulations. In fact, while approaching the weak coupling regime, autocorrelation effects prevent calculations from reaching very small values
of the lattice spacing, making continuum extrapolations and hence convergence to meaningful results difficult. We remark that in large scale lattice simulations, much more sophisticated definitions of the renormalized coupling $g_{\rm ren}$ are generally used \cite{Bruno:2017gxd, aoki2020flag}. These alternative definitions allow, besides other things, to disentangle cut-off effects, inherent to the plaquette coupling, from the true physical running. However, the alternative forms of $g_{\textrm{ren}}$ described in \cite{Bruno:2017gxd, aoki2020flag} are meaningful for large lattices only, and are hence beyond the capabilities of current quantum simulators. In this work, we allow for a proof-of-concept demonstration which paves the way to future improvements.\\
\\With our method (see Sec.~\ref{subsec:encodedHamiltonian_effectiveHamiltonianForPeriodicBoundaryConditions} and Ref.~\cite{paper1}), we can simulate the system both in the strong and in the weak coupling regimes, with very modest truncations. The most difficult region to simulate is characterized by $g^{-2} \simeq 1$, where convergence is studied in Ref.~\cite{paper1}. We demonstrate this in Fig.~\ref{fig:plaquettePeriodicPlot}, where the ground state expectation value of the plaquette operator $\expval{\Box}$ is plotted against $g^{-2}$. For PBC, the plaquette operator is given by $\Box = -\frac{g^2}{4} \hat{H}_{\textrm{B}}^{(\gamma)}$ where $\gamma = e$ ($\gamma = b$) refers to the electric (magnetic) representation [see Eqs.~\eqref{eq:periodicHamiltonianElectricB} and
\eqref{eq:periodicHamiltonianMagneticB}].
Accordingly, the blue and red lines are determined with these two representations, while the grey, dotted lines are obtained using perturbation theory, and describe well the exact results in the extremal regions. The left graph is for a truncation $l=1$ and the right for $l=2$.\\
%
%
%
\\In Sec.~\ref{sec:main_results}, we provide a protocol for quantum simulating
the system with a truncation $l = 1$ of the gauge field using 9 qubits.
As shown in Fig.~\ref{fig:plaquettePeriodicPlot}, using a truncation $l = 2$ around the region of the plot where $g^{-2} \simeq 1$ brings us much closer to the converged result. Using the same protocol as for $l = 1$ (see Sec.~\ref{subsec:trappedIonResults_periodicBoundaryConditions}),
the simulation for $l = 2$ requires $15$ qubits.
%
\section{Conclusions \& outlook}
\label{sec:conclusionsAndOutlook}
\noindent In this work, we proposed a protocol to observe 2D effects in LGTs on currently available quantum computers. By using the methods in Ref.~\cite{paper1}, we provided a practical VQE-based framework to simulate two toy models using NISQ devices. Importantly, we include the numerics for observing 2D phenomena in a basic building block of 2D QED with present-day quantum resources. \\
%
%
\\The effective models studied here include both dynamical matter and a non-minimal gauge field truncation, providing the novel opportunity to study several 2D effects in LGTs. More specifically, we showed how to observe dynamical generation of magnetic fields as a result of particle-antiparticle pair creation, and paved the way for an important first step towards simulating short distance quantities such as the running coupling of QED. While the protocols presented in Sec.~\ref{sec:main_results} are designed for trapped ion systems, our approach can be easily adapted to suit different types of quantum hardware.\\
\\One of the most appealing characteristics of our approach is that it can be generalized to more complex systems. Immediate extensions of our results include simulations of a 2D plane of multiple plaquettes, and of QED in three spatial dimensions. Moreover, including fermionic matter in the case of PBC can be done by following the same procedure as for OBC \cite{paper1}. It will also be interesting to explore implementations on non-qubit based hardware that is capable of representing gauge degrees of freedom with spin-$l$ systems ($l>1/2$) \cite{senko2015realization}. Another extension is the development of Trotter-type protocols to simulate real-time evolution
using the encoding given in Sec.~\ref{subsec:variationalQuantumSimulation_qubitEncoding}. As part of the quest to move towards quantum simulations of QCD, another possible extension is to progress from a U(1) gauge theory (QED) to a non-Abelian gauge theory, such as SU(2), and eventually to SU(3). Importantly, simulations of LGTs with larger lattice sizes will become feasible with future advancements in quantum computing, allowing for exciting possibilities, such as the ability to relate the running coupling of a gauge theory to a physical parameter and to make connections between quantum simulations and experiments in high energy physics. Ultimately, future quantum computers may offer the potential to also simulate models with a topological term, non-zero chemical potential, or real-time phenomena, effects which are very hard or even impossible to access with MCMC techniques.\\
\\The field of quantum simulations of LGTs is in its early exploratory stages and is rapidly developing. The type of problems considered here will act as test-bed for quantum computer implementations. By providing practical and experimentally feasible solutions for simulating gauge theories beyond 1D, our work opens up new pathways towards accessing regimes that are classically out of reach.
\section*{Acknowledgements}
\noindent We thank Rainer Blatt, Philipp Schindler, Thomas Monz, Raymond Laflamme, and Michele Mosca for fruitful and enlightening discussions, and Luca Masera for computational consulting.
This work has been supported by Transformative Quantum Technologies Program (CFREF), NSERC and the New Frontiers in Research Fund.
JFH acknowledges the Alexander von Humboldt Foundation in the form of a Feodor Lynen Fellowship.
CM acknowledges the Alfred P. Sloan foundation for a Sloan Research Fellowship.
AC acknowledges support from the Universitat Aut\'{o}noma de Barcelona Talent Research program,
from the Ministerio de Ciencia, Inovaci\'{o}n y Universidades (Contract No. FIS2017-86530-P),
from the the European Regional Development Fund (ERDF) within the ERDF Operational Program of Catalunya (project QUASICAT/QuantumCat),
and from the the European Union's Horizon 2020 research and innovation programme under the Grant Agreement No. 731473 (FWF QuantERA via QTFLAG I03769). Research in Innsbruck is supported by the European Union’s Horizon 2020 research and innovation programme under Grant Agreement No. 817482 (PASQuanS).\\
%
%
%
%
\appendix
\section{Effective Hamiltonian for a ladder of plaquettes with OBC}
\label{app:ElectricEnergyContribution}
\renewcommand{\arraystretch}{2}
\begin{table*}[t]
  \centering
  \caption{Electric field of each link in the ladder of Fig.~\ref{fig:openBoundaryConventions}(a). For the vertical outermost links, the cases of ``even" and ``odd" refer to the total number of plaquettes $N$.}
  \begin{ruledtabular}
  \begin{tabular}{lll} 
    Link type  & Even plaquette $n$             & Odd plaquette $k$ \\ \hline
    Horizontal independent  & $\hat{E}_{2n,2n+1}$           & $\hat{E}_{2k,2k+1}$  \\ 
    Horizontal dependent    & $\hat{E}_{2n-1, 2n+2} = -\hat{E}_{2n,2n+1} - \sum_{i=1}^{2n}\hat{q}_{i}$          & $\hat{E}_{2k-1, 2k+2} = \hat{E}_{2n-1, 2n+2}$  \\ 
    Vertical outermost      & $\hat{E}_{12} = \hat{E}_{23} + \hat{q}_2,~\hat{E}_{2N+1, 2N+2} = \hat{E}_{2N,2N+1} - \hat{q}_{2N+1}$
            &   $\hat{E}_{12} = \hat{E}_{23} + \hat{q}_2,~\hat{E}_{2N+2,2N+1} = -\hat{E}_{2N+1,2N+2}$ \\ 
    Vertical internal       & $\hat{E}_{2n+1,2n+2} = \hat{E}_{2n, 2n+1} + \hat{E}_{2n+2, 2n+3} ~+\sum_{i = 1}^{2n} \hat{q}_i - \hat{q}_{2n+2}$ & $\hat{E}_{2k+2,2k+1} = -\hat{E}_{2n+1, 2n+2}$   \\ 
  \end{tabular}
  \end{ruledtabular}
  \label{tab:electricTermDerivation}
\end{table*}

%
\noindent In this appendix, we derive an expression for the electric term in the Hamiltonian in Eq.~\eqref{eq:HtotOpenBoundary} for a ladder of $N$ plaquettes [see Fig.~\ref{fig:openBoundaryConventions}(a)]. Following the procedure outlined in Sec.~\ref{subsec:encodedHamiltonian_effectiveHamiltonianForOpenBoundaryConditions}, we implement Gauss' law to remove redundant gauge degrees of freedom and obtain an expression with only the independent gauge fields remaining. We then use the Jordan-Wigner transformation and the mapping in Eqs.~\eqref{eq:SpinTruncation} to arrive at the final Hamiltonian for the ladder. The magnetic, mass, and kinetic terms remain unchanged, except for the identity operators where the degrees of freedom are removed.\\
\\To write the electric term $\hat{H}_{\textrm{E}}$ in Eqs.~\eqref{eq:HtotOpenBoundary}, we express the electric field on each link in terms of the chosen independent gauge and matter degrees of freedom (see Sec.~\ref{subsec:encodedHamiltonian_LatticeQEDIn2+1Dimensions}). Categorizing links according to their orientation and location [defined in Fig.~\ref{fig:openBoundaryConventions}(a)], we examine the following cases: horizontal independent links, horizontal dependent links, outermost vertical links, and internal vertical links. Listed in Table~\ref{tab:electricTermDerivation} are the electric field contributions of each of these terms for even- and odd-numbered plaquettes, written in terms of the horizontal independent links.\\
\\Squaring each contribution in Table~\ref{tab:electricTermDerivation} and summing over the plaquettes allows us to arrive at an expression for the electric term in terms of only the independent gauge fields. Using the Jordan-Wigner transformation and Eqs.~\eqref{eq:SpinTruncation}-\eqref{eq:S-operator}, the Hamiltonian for the ladder can then be written
\begin{widetext}
\begin{subequations} \label{eq:OpenBoundarySpinHamiltonian}
  \begin{align}
    \hat{H} = &~\hat{H}_{\textrm{E}} + \hat{H}_{\textrm{B}} + \hat{H}_{\textrm{m}} + \hat{H}_{\textrm{kin}}, \\
    \hat{H}_{\textrm{E}} = &~\frac{g^2}{2}\Big\{\sum_{n=1}^N \Big[ \Big(\hat{S}^z_{2n,2n+1}\Big)^2 + \Big(\hat{S}^z_{2n,2n+1} + \sum_{i=1}^{2n}\hat{q}_{i}\Big)^2\Big] + \sum_{n = 1}^{N-1}\Big(\hat{S}^z_{2n, 2n+1} + \hat{S}^z_{2n+2, 2n+3} + \sum_{i=1}^{2n}\hat{q}_i + \hat{q}_{2n+2}\Big)^2  \nonumber \\
    &+ \Big(\hat{S}^z_{23}+ \hat{q}_2\Big)^2 + \Big(\hat{S}^z_{2N, 2N+1} - \hat{q}_{2N+1}\Big)^2 \Big\} \\
    \hat{H}_{\textrm{B}} = &-\frac{1}{2g^2}\sum_{n = 1}^{N} \Big(\hat{V}^-_{2n,2n+1} + \hat{V}^{+}_{2n, 2n+1}\Big), \\
    \hat{H}_{\textrm{m}} = &~\frac{m}{2} \sum_{i = 1}^{2N+2} (-1)^{i+1} \hat{\sigma}_i^z, \\
    \hat{H}_{\textrm{kin}} = &-i\Omega \Big\{\hat{\sigma}_1^+ \hat{\sigma}_2^- + \sum_{n=1}^N \Big[\hat{\sigma}_{2n}^- \hat{V}^+_{2n, 2n+1} \hat{\sigma}_{2n+1}^+ + (-1)^n(\hat{\sigma}_{2n+1}^+\hat{\sigma}_{2n+2}^-) -  \hat{\sigma}_{2n-1}^+ \hat{\sigma}_{2n}^{z} \hat{\sigma}_{2n+1}^{z} \hat{\sigma}_{2n+2}^-\Big]\Big\} + \textrm{H.c.},\label{eq:OpenBoundarySpinHamiltoniankin}
  \end{align}
\end{subequations}
\end{widetext}
where $\hat{\sigma}_i^{\pm} = \frac{1}{2}(\hat{\sigma}_i^x \pm \hat{\sigma}_i^y)$ and $\hat{\sigma}_i^x, \hat{\sigma}_i^y, \hat{\sigma}_i^z$ are the Pauli operators acting on the $i^{\textrm{th}}$ qubit. Here, the charge operator $\hat{q}_i$ under the Jordan-Wigner transformation becomes $\hat{q}_i = \frac{Q}{2}\big(\hat{\sigma}_i^z + (-1)^{i+1}\big)$.
\section{Optimal partitioning for Hamiltonian averaging}
\label{app:optimalPartitioning}
\noindent Our VQE algorithm requires the measurement of the energy of the variational state, i.e., the cost function $\mathcal{C}(\boldsymbol{\theta})$. Precise estimation of the expectation value of a complicated Hamiltonian demands, in general, a large number of measurements, which translates to extensive computational runtime. In this section, we show how the number of measurements required for the proposed quantum simulations can be substantially reduced by combining the standard approach \cite{peruzzo2014variational, mcclean2016theory} with the measurement strategy put forward in Ref.~\cite{jena2019pauli}. We adapt the latter to our models and provide circuits for their implementation on different types of quantum hardware. The described scheme assumes local measurements only.\\
\\By using Pauli operators as a basis, we can express any target Hamiltonian $\hat{H}_T$ as
\begin{subequations}
\begin{align}
	\hat{H}_T = &\sum_{k = 1}^{K} c_k \hat{Q}_k, \label{eq:hamiltoniandecomposition}\\
	\hat{Q}_k = &\bigotimes_{i=1}^{n} \hat{\sigma}_{i}^{k_i},
\end{align}
\end{subequations}
where $K$ is the number of terms in the decomposition, $n$ is the dimension of the system, $\hat{\sigma}_{i}^{k_i}$ is a Pauli operator acting on the $i^{\textrm{th}}$ qubit and $k_i$ can either be $0$ (identity), $x$, $y$, or $z$ for any $i$. We remark that, for our qubit encoding (see Sec.~\ref{subsec:variationalQuantumSimulation_qubitEncoding}), decomposing the Hamiltonian as in Eq.~\eqref{eq:hamiltoniandecomposition} can be efficiently done using the relations given in Eq.~\eqref{eq:encodedoperators}.\\
\\The cost function $\mathcal{C}(\boldsymbol{\theta})$ is evaluated by measuring the expectation values $\expval{\hat{Q}_{k}}$ with single-qubit Pauli measurements and calculating the weighted sum $\expval{\hat{H}_T} = \sum_{k=1}^{K} c_k \expval{\hat{Q}_k}$. Operators $\hat{Q}_k$ that commute with each other can be measured simultaneously and will be grouped together in the following. Thus, we rewrite Eq.~\eqref{eq:hamiltoniandecomposition} as $\hat{H}_T = \sum_{m = 1}^{M} \hat{R}_m$, where
\begin{align}
	\hat{R}_m = \sum_{k \in \mathcal{V}_m} c_k \hat{Q}_k. \label{eq:Rm}
\end{align}
Here, the set $\mathcal{V}_m$ contains indices of commuting operators $\hat{Q}_k$. Importantly, the number $M$ of commuting sets can be drastically smaller than the total number of Pauli operators $K$ in Eq.~\eqref{eq:hamiltoniandecomposition}, leading to a reduction in the experimental runtime.\\
\\To successfully run a VQE algorithm, we must be able to calculate the average values of the energy for any variational state preparation. Due to the statistical nature of quantum mechanics, the exact average value of any observable is unknown. Thus, we need to approximate the energy with an estimator. For a generic operator $\hat{O}$, we denote the estimator of its expectation value $\expval{\hat{O}}$ with $\bar{O}$. Depending on the intrinsic variance $(\Delta \hat{O})^2$ and the number $N$ of repeated measurements performed, the estimator $\bar{O}$ is affected by a statistical error, which is characterized by its variance $\textrm{Var}[\bar{O}]$. \\
\\To approximate the system's energy, we choose beforehand a certain precision, which is generally problem and/or application dependent. In practice, this means that we are required to fix a threshold $\epsilon^2$, and repeat energy measurements $\bar{H}_T$ until the variance $\textrm{Var}[\bar{H}_T]$ becomes smaller than $\epsilon^2$. Since $\hat{H}_T$ is the sum of the independent operators $\hat{R}_m$ [see Eq.~\eqref{eq:Rm}], $\bar{H}_T$ can be expressed in terms of the estimators for the $\expval{\hat{R}_m}$ as $\bar{H}_T = \sum_{m = 1}^{M} \bar{R}_m$. More specifically, each estimator $\bar{R}_m$ is found by averaging $N_m$ measurement outcomes $\{r_{m}^{i}\}$, i.e.,
\begin{align}
	\bar{R}_m = & \frac{1}{N_m} \sum_{i = 1}^{N_m} r_m^i.
\end{align}
For sufficiently large $N_m$, the variance of the estimators $\bar{R}_m$ is $\textrm{Var}[\bar{R}_m] \simeq (\Delta \hat{R}_m)^2/N_m$, where $(\Delta \hat{R}_m)^2$ is the intrinsic variance of the operator $\hat{R}_m$ with respect to the considered state. Since the operators $\hat{R}_m$ are measured in different experimental runs, the estimators $\bar{R}_m$ are independent and thus have zero covariance $\textrm{Cov}[\bar{R}_i, \bar{R}_j]=0$. As such, we can express $\textrm{Var}[\bar{H}_T]$ as the sum
\begin{align}
	\textrm{Var}[\bar{H}_T] = &\sum_{m=1}^{M} \textrm{Var}[ \bar{R}_m].
\end{align}
Achieving an error in $\bar{H}_T$ of no more than $\epsilon$ requires $\textrm{Var}[\bar{H}_T] = \epsilon^2$. Assuming equal contribution from each of the $\bar{R}_m$, we write $\textrm{Var}[\bar{R}_m] = \epsilon^2/M$. Substituting $\textrm{Var}[\bar{R}_m] \simeq (\Delta \hat{R}_m)^2/N_m$, the number of measurements for each $m$ is $N_m \simeq M(\Delta \hat{R}_m)^2/\epsilon^2$, and, summing over $m$, the total number of measurements $N$ is
\begin{align}
	N \simeq &~\frac{M}{\epsilon^2} \sum_{m = 1}^{M} (\Delta \hat{R}_m)^2. \label{eq:numofmeasurements}
\end{align}
\\Besides the chosen threshold $\epsilon^2$, the total number of measurements $N$ is proportional to the intrinsic variances $(\Delta \hat{R}_m)^2$ and the number of partitioned sets $M$. In general, there are many different ways to partition the $\hat{Q}_k$ into the $\hat{R}_m$, and we seek to identify a strategy which minimizes both $M$ and $(\Delta \hat{R}_m)^2$. Importantly, $(\Delta \hat{R}_m)^2$ depends on the intrinsic variances and the covariances of the $\hat{Q}_k$ from which each $\hat{R}_m$ is comprised. While a particular partitioning strategy might achieve $M < K$, it might also group together $\hat{Q}_k$ with positive covariance, thereby adding to $(\Delta\hat{R}_m)^2$ and increasing the number of measurements needed to achieve $\epsilon^2$ precision of $\bar{H}_T$. For instance, as pointed out in Ref.~\cite{mcclean2016theory}, the Hamiltonian
\begin{align}
	H = & -(\hat{\sigma}_{1}^{x} \hat{\sigma}_{2}^{x} + \hat{\sigma}_{1}^{y} \hat{\sigma}_{2}^{y}) + \hat{\sigma}_{1}^{z} \hat{\sigma}_{2}^{z} + \hat{\sigma}_{1}^{z} + \hat{\sigma}_{2}^{z}
\end{align}
could be partitioned in the following ways:
\begin{enumerate}
	\item $\{-\hat{\sigma}_{1}^{x} \hat{\sigma}_{2}^{x}\}, \{-\hat{\sigma}_{1}^{y} \hat{\sigma}_{2}^{y}\}, \{\hat{\sigma}_{1}^{z} \hat{\sigma}_{2}^{z}\}, \{\hat{\sigma}_{1}^{z}\}, \{\hat{\sigma}_{2}^{z}\},$
	\item $\{-\hat{\sigma}_{1}^{x} \hat{\sigma}_{2}^{x}\}, \{-\hat{\sigma}_{1}^{y} \hat{\sigma}_{2}^{y}, ~\hat{\sigma}_{1}^{z} \hat{\sigma}_{2}^{z}\}, \{\hat{\sigma}_{1}^{z}, ~\hat{\sigma}_{2}^{z}\},$
	\item $\{-\hat{\sigma}_{1}^{x} \hat{\sigma}_{2}^{x}, ~-\hat{\sigma}_{1}^{y} \hat{\sigma}_{2}^{y}, ~\hat{\sigma}_{1}^{z} \hat{\sigma}_{2}^{z}\}, \{\hat{\sigma}_{1}^{z}, ~\hat{\sigma}_{2}^{z}\}.$
\end{enumerate}
When measuring the state $\ket{\Psi} = \ket{0 1}$, partition $1$ has variance $\sum_{m=1}^{5}(\Delta\hat{R}_m)^2 = 2$ and $M=5$ commuting sets for a total of $N^{(1)} = 10/\epsilon^2$ measurements [see Eq.~\eqref{eq:numofmeasurements}]. Partition $2$ has the same variance because the operators within the sets have zero covariance, but here $M=3$, so $N^{(2)} = 6/\epsilon^2$. Although partition $3$ has only $M = 2$ commuting sets, the covariance between $\hat{\sigma}_{1}^{x} \hat{\sigma}_{2}^{x}$ and $\hat{\sigma}_{1}^{y} \hat{\sigma}_{2}^{y}$ adds to $(\Delta \hat{R}_1)^2$, giving $N^{(3)} = 8/\epsilon^2$ measurements. In this example, partition $2$ gives the fewest number of measurements even though partition $3$ has the fewest number of commuting sets.\\
\begin{figure}
	\centering
	\includegraphics[width=\columnwidth]{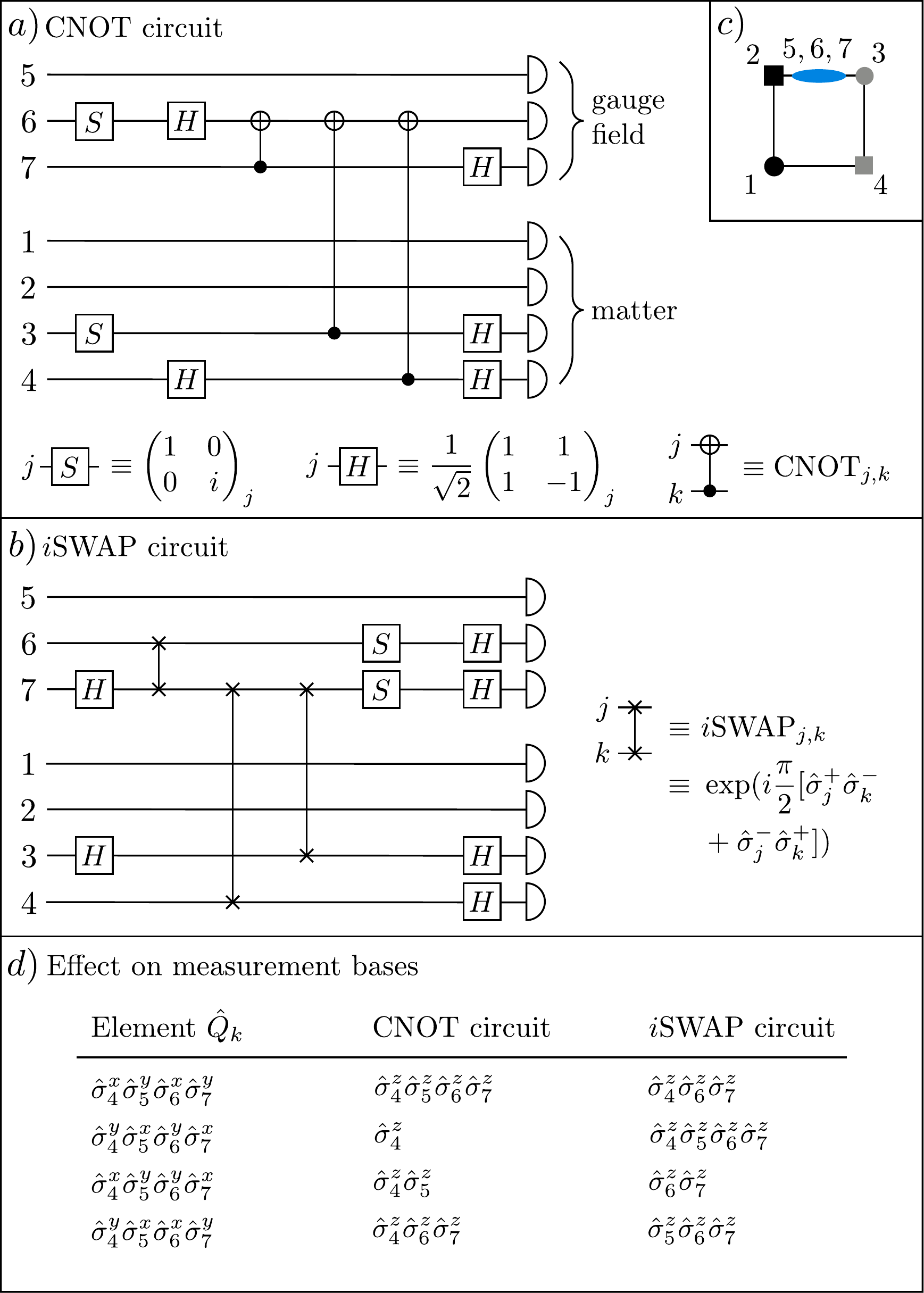}
	\caption{An example of the measurement protocol explained in the main text and App.~\ref{app:optimalPartitioning}, applied to the open boundary plaquette (see Sec.~\ref{subsec:encodedHamiltonian_effectiveHamiltonianForOpenBoundaryConditions}). The shown circuits diagonalize a measurements set $\hat{R}_m$ using CNOT gates (a) and $i$SWAP gates (b) as an entangling operation. The qubit labelling conventions are shown in (c). Applying the circuits before performing local measurements in a VQE protocol amounts to effectively performing measurements in an entangled basis. The table in (d) lists the elements $\hat{Q}_{k}$ in the chosen measurement set and their corresponding diagonalized elements using the CNOT circuit and the $i$SWAP circuit.}
	\label{fig:measurementCircuit}
\end{figure}
\\The variances and covariances of $\hat{Q}_k$ are dependent on the state $\ket{\Psi}$ being measured, which is generally unknown. Thus, to reduce the total number of measurements required, we seek a partitioning strategy that yields the minimum number of commuting sets, i.e., we minimize $M$ without regard for the variance of the partitions. In principle, it is possible to design an algorithm which, while acquiring data, reorganizes the partitions in order to minimize their variances, but this goes beyond the scope of this work \cite{jena2019decomposition}.\\
\\The algorithm presented in Ref.~\cite{jena2019pauli} finds the partition of the $\hat{Q}_k$ into $\hat{R}_m$ that gives the near-minimum number of commuting sets $M$ \footnote{This is a greedy algorithm \cite{cormen2009introduction} that scales well with the size of the system and the number of terms in the decomposition, but it is not guaranteed to find the optimal solution. It would be possible to develop an algorithm that finds the absolute minimum of $M$ at the expense of additional computational resources.}. Although commuting operators can in principle be measured simultaneously, from an experimental point of view, only those operators which bitwise commute can be easily measured simultaneously with the local measurements provided by the quantum device. To exemplify this, consider $\hat{\sigma}_{1}^{x} \hat{\sigma}_{2}^{x}$ and $\hat{\sigma}_{1}^{z} \hat{\sigma}_{2}^{z}$. While these operators commute and thus can be simultaneously measured, there are no single qubit operations which rotate them into a diagonal form. To achieve this, two-qubit gates are required. Ref.~\cite{jena2019pauli} provides the procedure for building a quantum circuit which diagonalizes the Pauli operators $\hat{Q}_k$ within a commuting set (for an example, see Fig.~\ref{fig:measurementCircuit}). This circuit is to be placed after the VQE circuit and before the local measurements, effectively rotating the measurement basis and allowing us to measure in a complicated entangled basis. As such, all commuting operators within a commuting set, bitwise or not, can be measured simultaneously.\\
\\The gate set from which the quantum circuits are built can be adapted to suit the quantum computing platform, using for instance CNOT operations as entangling gates for superconducting platforms and $i$SWAP gates for ion-based platforms. While it is possible to express a CNOT gate in terms of $i$SWAP gates, and vice versa, this leads to linear overhead in circuit depth, which impacts the feasibility of experiments with NISQ devices. Therefore, we adapt the algorithm in Ref.~\cite{jena2019pauli} according to the preferred type of two-qubit gate to avoid this linear overhead in the number of gates. As an example, Fig.~\ref{fig:measurementCircuit}(a) and \ref{fig:measurementCircuit}(b) show two possible circuits which diagonalize one of the partitions $\hat{R}_m$ of the plaquette with OBC introduced in Sec.~\ref{subsec:encodedHamiltonian_effectiveHamiltonianForOpenBoundaryConditions}, with the qubit labelling convention shown in panel (c). The elements $\hat{Q}_k$ of this partition are shown in Fig.~\ref{fig:measurementCircuit}(d), along with the diagonal forms obtained after applying the CNOT and $i$SWAP circuits (for details see Refs.~\cite{jena2019pauli, jena2019decomposition}).\\
\\In the worst-case scenario, the depth of the additional circuits scales quadratically with $K$, however, in practice, we often see linear scaling. The computing time of the algorithm in Ref.~\cite{jena2019pauli} also scales quadratically in $K$, while the number of qubits in the system does not play a significant role in the complexity \footnote{Although an increasing number of qubits brings the possibility of exponentially many decomposition terms $K$, in many physical scenarios, $K$ is limited by the symmetries of the system, and we do not see exponential dependence.}. By applying this method to the Hamiltonian of Eqs.~\eqref{eq:singlePlaquetteHamiltonianEncoded}, we reduce the number of measurement sets (i.e., commuting sets) from $49$ down to $M = 5$ in the case of OBC. With PBC, the number of measurement sets is reduced from $295$ to $M = 11$ for the electric representation [Hamiltonian in Eqs.~\eqref{eq:periodicHamiltonianElectricEncoded}] and from $138$ down to $M = 15$ for the magnetic basis [Hamiltonian in Eqs.~\eqref{eq:periodicHamiltonianMagneticEncoded}]. The depth of the circuits used to implement the measurements is around ten in the worst-case scenario.\\
\section{Perturbation theory}
\label{app:PerturbationTheory}
\noindent In this appendix, we outline the perturbative algorithm used to determine the analytical results presented in Sec.~\ref{subsec:quantumSimulation_openBoundaryConditions}.\\
\\We consider the system Hamiltonian to be $\hat{H} = \hat{H}_0 + \hat{H}_{\textrm{kin}}$, where $\hat{H}_0 = \hat{H}_{\textrm{E}} + \hat{H}_{\textrm{m}}$ [see Eqs~\eqref{eq:singlePlaquetteHamiltonianHE} and \eqref{eq:singlePlaquetteHamiltonianHm}] is the bare Hamiltonian and $\hat{H}_{\textrm{kin}}$ [see Eq.~\eqref{eq:singlePlaquetteHamiltonianHkin}] is the perturbation. Following standard perturbation theory, the corrected state and energy for eigenvalue $i$ up to order $n$ can be written as
\begin{subequations} \label{eq:perturbativeCorrections}
\begin{align}
  \ket{\Psi_{i}} = &\sum_{j = 0}^n  \ket{\Psi_{i}^{(j)}},\\
  E_{i} = &\sum_{j = 0}^n E_{i}^{(j)}.
\end{align}
\end{subequations}
The unperturbed state $\ket{\Psi_{i}^{(0)}}$ can be any of the eigenstates of $\hat{H}_0$. Our goal is to find the corrections for $j \geq 1$ in Eqs.~\eqref{eq:perturbativeCorrections} up to an arbitrary order and for any initial state. We require higher orders in perturbation theory because the leading and the second leading contributions are found at fourth and sixth orders, respectively. Furthermore, degeneracies in our system are often lifted at fourth order. Importantly, here we do not truncate the gauge field operators. Since the action of $\hat{H}_{\textrm{kin}}$ is of creating and/or annihilating, whenever possible, a fermion-antifermion pair and fix the gauge field between, we can select all contributing states at each perturbation order.\\
\begin{figure*}[t]
  \includegraphics[width=1.3\columnwidth]{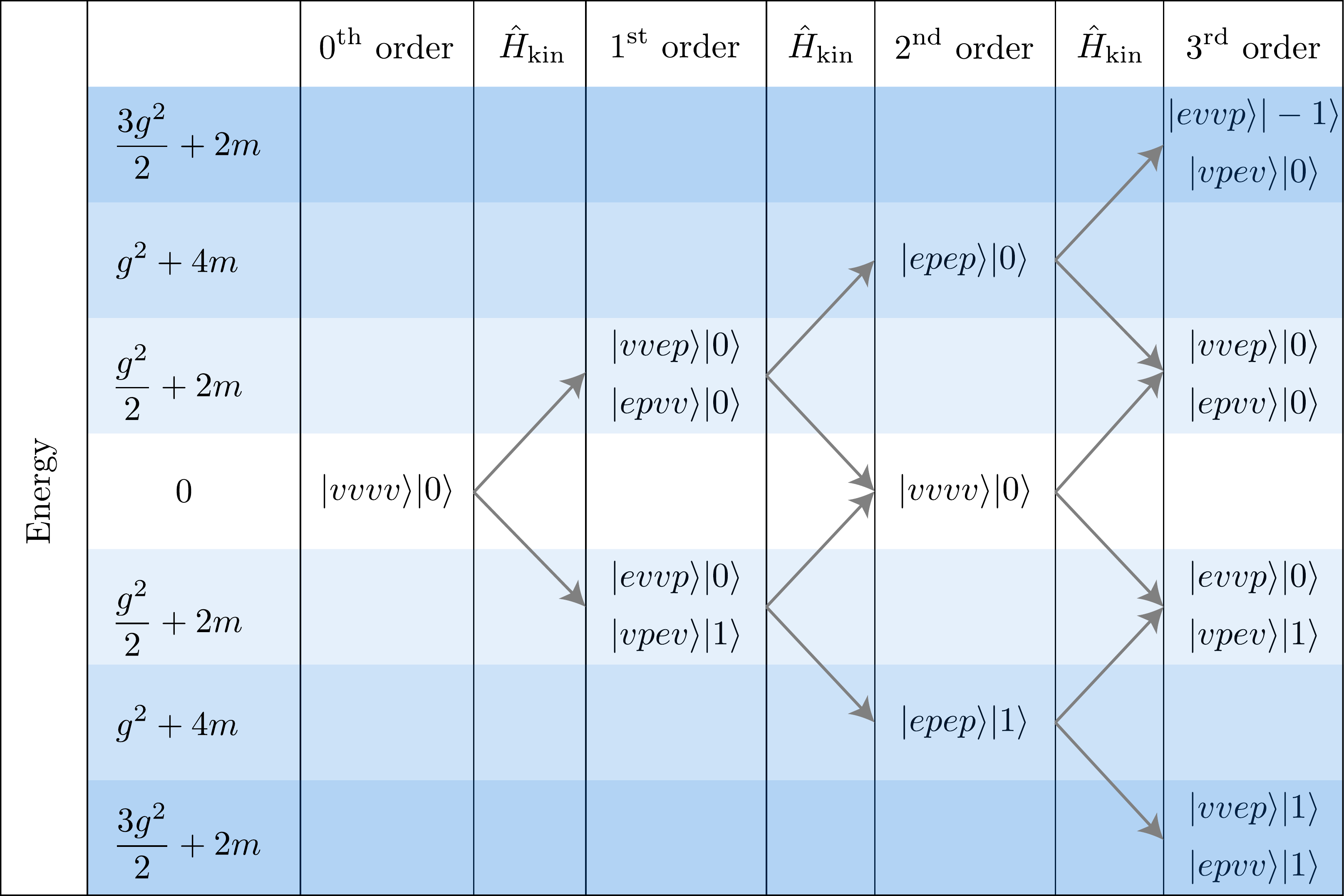}
  \caption{Schematic description of our perturbative algorithm when starting from the vacuum. In the first column, we show the bare energies of the corresponding states. Each application of the perturbation $\hat{H}_{\textrm{kin}}$ allows for creating and annihilating particle-antiparticle pairs, as depicted by the tree structure. At each iteration, our algorithm expands the Hilbert space to include new terms contributing to the correction. This allows for efficiently using the resources, ultimately reaching several hundred orders in perturbation theory.}
  \label{fig:treeFig}
\end{figure*}
\\The eigenstates and corresponding eigenvalues of the bare Hamiltonian $\hat{H}_0$ are denoted as $\ket{\Psi_{i,\mu}^{(0)}}$ and $E_{i}^{(0)}$, where $\mu$ is an index used within the degenerate subspace that contains the unperturbed state $\ket{\Psi_{i}^{(0)}}$. As an example, the six lowest energy eigenstates of $\hat{H}_0$ are as follows (see Fig.~\ref{fig:treeFig}): the non-degenerate vacuum state $\ket{vvvv}\ket{0}$ with null energy; the four-fold degenerate two-particle states $\ket{epvv}\ket{0}$, $\ket{vpev}\ket{1}$, $\ket{vvep}\ket{0}$, and $\ket{evvp}\ket{0}$, each with energy $\frac{g^2}{2} + 2m$; and the two-fold degenerate fully-filled plaquette states $\ket{epep}\ket{0}$ and $\ket{epep}\ket{1}$, each with energy $g^2 + 4m$ [see Fig.~\ref{fig:openBoundaryConventions}(e)].\\
\\Acting on an eigenstate of $\hat{H}_0$ with the kinetic term creates and annihilates particle-antiparticle pairs, and fixes the field on the links to satisfy Gauss' law. Successive applications of $\hat{H}_{\textrm{kin}}$ on the vacuum are shown in Fig.~\ref{fig:treeFig}. This illustration shows that acting $\hat{H}_{\textrm{kin}}$ on a state an odd number of times results in a state that is orthogonal to the initial state, implying that odd order perturbative energy corrections are vanishing [see Eq.~\eqref{eq:perturbativeEnergyCorr}].\\
\\At the beginning of our algorithm, the initial state $\ket{\Psi_{i}^{(0)}}$ is set to an unknown superposition of states within the selected eigensubspace $i$, i.e., $\ket{\Psi_{i}^{(0)}} = \sum_{\nu} d_{\nu}^{(0)}\ket{\Psi_{i, \nu}^{(0)}}$. If the subspace $i$ is degenerate, the coefficients $d_{\nu}^{(0)}$ are determined by lifting -- when possible -- the degeneracy at higher orders (see below). The $n^{th}$ order perturbative corrections are calculated as
\begin{subequations} \label{eq:perturbativeCorrections2}
  \begin{align}
    \braket{\Psi_k^{(0)}}{\Psi_{i}^{(n)}} = &~\frac{\bra{\Psi_k^{(0)}} \hat{H}_{\rm kin} \ket{\Psi_{i}^{(n-1)}}}{E_i^{(0)} - E_k^{(0)}} \nonumber\\
    &- \sum_{j=1}^{n-1} E_{i}^{(j)} \frac{\braket{\Psi_k^{(0)}}{\Psi_{i}^{(n-j)}}}{E_i^{(0)}-E_k^{(0)}},\\
    E_{i}^{(n)} = &~\bra{\Psi_{i}^{(0)}} \hat{H}_{\rm kin} \ket{\Psi_{i}^{(n-1)}} \nonumber\\&- \sum_{j=1}^{n-1} E_{i}^{(j)}\braket{\Psi_{i}^{(j)}}{\Psi_{i}^{(n-j)}}, \label{eq:perturbativeEnergyCorr} \\
    \braket{\Psi_{i, \nu}^{(0)}}{\Psi_{i}^{(n)}} = &~d_{\nu}^{(n)},
  \end{align}
\end{subequations}
where $\ket{\Psi_k^{(0)}}$ with energy $E_{k}^{(0)}$ are the eigenstates of $\hat{H}_0$ outside the degenerate eigensubspace which contains the initial state. The overlap $d_{\nu}^{(n)}$ of the state correction with the eigenstates $\ket{\Psi_{i, \nu}^{(0)}}$ within the same degenerate subspace can be determined by imposing normalization when the degeneracy is lifted. The specific form of Eqs.~\eqref{eq:perturbativeCorrections2} follows from the symmetries that our perturbation $H_{\textrm{kin}}$ satisfies. In particular, $E_{i}^{(n)} = d_{\nu}^{(n)} = 0$ for odd $n$.\\
\\To lift a degeneracy, the coefficients $d_{\nu}^{(0)}$ need to be found by diagonalizing the matrix equation
\begin{align}
  \bra{\Psi_{i,\nu}^{(0)}} \hat{H}_{\rm kin} \ket{\Psi_{i}^{(n-1)}} = \sum_{j=2}^{n} E_{i}^{(j)} \braket{\Psi_{i,\nu}^{(0)}}{\Psi_{i}^{(n-j)}}. \label{eq:solveDegeneracy}
\end{align}
Eq.~\eqref{eq:solveDegeneracy} can only be solved at the order $m$ at which the perturbation lifts the degeneracy. For instance, when starting from the fully-filled plaquette states $\ket{epep}\ket{0}$ and $\ket{epep}\ket{1}$, the two-fold degeneracy is lifted at $m=4$. However, the four-fold degeneracy between the two-particle states is only partially lifted at $m=2$ into two two-fold degenerate systems $\{\ket{evvp}\ket{0}, \ket{vpev}\ket{1}\}$ and $\{\ket{epvv}\ket{0}, \ket{vvep}\ket{0}\}$. The former is fully lifted at $m=4$. The latter remains degenerate for all orders because these states satisfy the same symmetry as the perturbation $\hat{H}_{\textrm{kin}}$.\\
\\As mentioned in Sec.~\ref{subsec:quantumSimulation_openBoundaryConditions}, the jump coordinate $g_{\textrm{c}}$ can be estimated using perturbation theory. The jump occurs when the ground state transitions from $\ket{f_-}$ to $\ket{f_+}$ (see Sec.~\ref{subsec:quantumSimulation_openBoundaryConditions}), i.e., when the energies of the two states are equal. Given that the bare energies $E_{i}^{(0)}$ are the same, this means that for $g = g_{\textrm{c}}$ the sum of the kinetic $E_{\textrm{kin}}$ and magnetic $E_{\textrm{B}} = - \expval{\Box}/(2g^2)$ energies for the two states are equal. While $E_{\textrm{kin}}$ is the energy correction given from perturbation theory, one can determine $E_{\textrm{B}}$ using the corrected state $\ket{\Psi_i} = \ket{f_{\pm}}$. Thus, we can write $E_{\textrm{B}}^{(+)} + E_{\textrm{kin}}^{(+)} = E_{\textrm{B}}^{(-)} + E_{\textrm{kin}}^{(-)}$, where the superscript $(\pm)$ indicates the perturbative state $\ket{f_{\pm}}$ used. By solving for $g$, we arrive at the estimated jump coordinate up to order $n$ in perturbation theory.\\
\section{Ground state of open boundary plaquette in the weak coupling limit}
\label{app:newAppendix}
\noindent In this appendix, we derive the ground state of the plaquette with OBC in the weak coupling regime of $g^{-2} \gg 1$. The following derivations are valid for the non-perturbative regime described in Sec.~\ref{subsec:quantumSimulation_openBoundaryConditions} [see Fig.~\ref{fig:plaquetteNonPerturbative}(d)], where $\Omega \gg \lvert m \rvert$ and $\Omega \ll g^{-2}$. The Hamiltonian is given in Eqs.~\eqref{eq:HtotOpenBoundary}.\\
\\For large $g^{-2}$, the Hamiltonian is dominated by the magnetic term $\hat{H}_{\textrm{B}}$, which contains only gauge field operators. As such, the ground state is required to be a product state between some matter configuration and ground state of the magnetic term. This is confirmed by the plot of the entanglement entropy in Fig.~\ref{fig:plaquetteNonPerturbative}(a), which shows that the ground state is separable between matter and gauge components for large $g^{-2}$. Therefore, we set the ground state to be a tensor product $\ket{\Psi} = \ket{\psi}\ket{\textrm{GS}^{(b)}}$ between some matter configuration $\ket{\psi} = \sum_{j=1}^6 c_j \ket{m_j}$ and the ground state of the magnetic term $\ket{\textrm{GS}^{(b)}}$. Here, $\ket{m_j}$ are the allowed matter configurations $\{\ket{vvvv}$, $\ket{epvv}$, $\ket{vpev}$, $\ket{vvep}$, $\ket{evvp}$, $\ket{epep}\}$ for $j = 1,~2,\dots,~6$, respectively (see Table~\ref{tab:optimalcj}). While the magnetic term is dominant, it does not influence the matter configuration. For simplicity, we set the mass $m = 0$ (the following procedure can be generalized for non-zero $m$), leaving the remaining candidates for determination of the matter state to be $\hat{H}_{\textrm{E}}$ and $\hat{H}_{\textrm{kin}}$. We are in the limit $\Omega \gg 1$, $g^{-2} \gg 1$, and so the kinetic term is dominant over the electric term. By minimizing $\bra{\Psi} \hat{H}_{\textrm{kin}} \ket{\Psi}$ over the $c_j$, we can estimate the matter component of the ground state at large $g^{-2}$.\\
\\The expression for $\bra{\Psi} \hat{H}_{\textrm{kin}} \ket{\Psi}$ can be written as
\begin{align}
  \bra{\Psi} \hat{H}_{\textrm{kin}} \ket{\Psi} = &~i\Omega \big[ c_1^* c_2 + c_1^* c_5 + c_3^* c_6 \nonumber\\& + c_4^* c_1 + c_4^* c_6 + c_6^* c_2 \nonumber\\&+ \big(c_1^* c_3 + c_5^* c_6 \big) \bra{\textrm{GS}^{(b)}} \hat{V}^{-}   \ket{\textrm{GS}^{(b)}} \big] \nonumber\\&+ \textrm{H.c.}, \label{eq:optimalcj}
\end{align}
and the optimal $c_j$ are shown in Table~\ref{tab:optimalcj} for truncation $l = 1$ and $l \rightarrow \infty$. The residual probabilities of the states $\ket{vpev}$ and $\ket{evvp}$ in Fig.~\ref{fig:plaquetteNonPerturbative}(d) are an effect of the $l = 1$ truncation.
\begin{table}[]
	\centering
	\caption{The $c_j$ for different truncations obtained from minimizing Eq.~\eqref{eq:optimalcj} which correspond to the matter component of the ground state of the plaquette with OBC Hamiltonian for large $g^{-2}$.}
\begin{ruledtabular}
\begin{tabular}{llll}
$j$ & $\ket{m_j}$ & $c_j$ ($l = 1$) & $c_j$ ($l \rightarrow \infty$) \\ \hline
1   & $\ket{vvvv}$                     & $-0.50i$    &  $-i/2$  \\ 
2   & $\ket{epvv}$                     & $0.49$      &  $1/2$  \\ 
3   & $\ket{vpev}$                     & $-0.073$     &  $0$  \\ 
4   & $\ket{vvep}$                     & $-0.49$      &  $-1/2$  \\ 
5   & $\ket{evvp}$                     & $0.073$     &  $0$  \\ 
6   & $\ket{epep}$                     & $-0.50i$ 		 & 	$-i/2$	\\ 
\end{tabular}
\end{ruledtabular}
\label{tab:optimalcj}
\end{table}
Constructing the estimated ground state $\ket{\Psi} = \ket{\psi}\ket{\textrm{GS}^{(b)}}$ using the optimal $c_j$, we find that $\ket{\Psi}$ has over $99.9\%$ overlap with the ground state obtained via exact diagonalization of the Hamiltonian in Eqs.~\eqref{eq:HtotOpenBoundary} for large $g^{-2}$.
%


\end{document}